\documentclass{article}

\usepackage{arxiv}

\usepackage[utf8]{inputenc} % allow utf-8 input
\usepackage[T1]{fontenc}    % use 8-bit T1 fonts
\usepackage{hyperref}       % hyperlinks
\usepackage{url}            % simple URL typesetting
\usepackage{booktabs}       % professional-quality tables
\usepackage{amsfonts}       % blackboard math symbols
\usepackage{nicefrac}       % compact symbols for 1/2, etc.
\usepackage{microtype}      % microtypography
\usepackage{lipsum}		% Can be removed after putting your text content
\usepackage{graphicx}
\usepackage{natbib}
\usepackage{doi}

\usepackage{algorithm} 
\usepackage{algpseudocode}
\usepackage{ulem}
\usepackage{mathptmx}
\usepackage{amsmath}
\usepackage{xcolor}

\title{Seismic Modeling and Migration with Random Boundaries on the NEC SX-Aurora TSUBASA}

\author{ \href{https://orcid.org/0000-0002-1420-9118}{\includegraphics[scale=0.06]{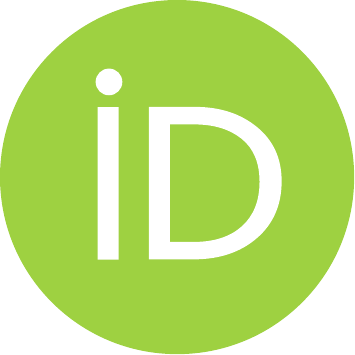}\hspace{1mm}Carlos H. S.~Barbosa} \\
	Dept. of Civil Engineering, COPPE\\
	Federal University of Rio de Janeiro\\
	Rio de Janeiro - Brazil \\
	\texttt{c.barbosa@nacad.ufrj.br} \\
	\And
	\href{https://orcid.org/0000-0002-4764-1142}{\includegraphics[scale=0.06]{orcid.pdf}\hspace{1mm}Alvaro L. G. A.~Coutinho} \\
	Dept. of Civil Engineering, COPPE\\
	Federal University of Rio de Janeiro\\
	Rio de Janeiro - Brazil \\
	\texttt{alvaro@nacad.ufrj.br} \\
}

\renewcommand{\shorttitle}{Seismic Modeling and Migration on the SX-Aurora TSUBASA}

\hypersetup{
pdftitle={Seismic Modeling and Migration with Random Boundaries on the NEC SX-Aurora TSUBASA and NVIDIA Volta V100},
pdfsubject={q-bio.NC, q-bio.QM},
pdfauthor={Carlos H. S.~Barbosa, Alvaro L. G. A.~Coutinho},
pdfkeywords={HPC, seismic modeling, RTM, OpenACC, vector processor},
}

\begin{document}
\maketitle

\begin{abstract}
Seismic imaging is a computationally demanding and data-intensive activity in the oil and gas industry. %However, the gap between computer processing speed and memory access has been growing for decades. 
Reverse Time Migration (RTM) used in seismic applications needs to store the forward-propagated wavefield (or source wavefield) on disk. Aiming to mitigate the storage demand, we develop an RTM that implements the source wavefield reconstruction by introducing a new wave equation to the problem. We adjust the initial and boundary conditions to take advantage of the properties of  random boundary conditions (RBC). The RBC does not suppress unwanted waves coming from the artificial boundary enabling the full wavefield recovery. Besides, it explores low correlations with non-coherent signals due to the random velocities in the boundary. We also develop compiler-guided implementations on a vector processor for seismic modeling and RTM, essential for Least-square Migration, Full Waveform Inversion, and Uncertainty Quantification applications. We test the seismic modeling and RTM on the 2-D Marmousi benchmark and 3-D HPC4E Seismic Test Suite. The numerical experiments show that the RTM which implements the wavefield reconstruction presents the best results in terms of execution time and hard disk demand. Lastly, the vector processor implementation is the one that requires fewer code modifications compared to the optimized baseline versions of the seismic modeling and RTM and GPU implementations, particularly for large 3D grids.
\end{abstract}

% keywords can be removed
\keywords{HPC \and seismic modeling \and RTM \and OpenACC \and vector processor}

\section{Introduction}

Reverse Time Migration (RTM) is a depth migration technique that provides a reliable high-resolution representation of the Earth subsurface useful for seismic interpretation, and reservoir characterization \citep{zhou2018reverse}. The RTM is based on the two-way wave equation and an appropriate imaging condition. Generally, the two-way wave equation is solved by numerical methods such as the Finite Difference Method (FDM) and the Finite Element Method (FEM). Besides, some imaging conditions need the computational implementation of the forward-propagated wavefield (or source wavefield) for further access in reversal order to build the seismic image.

Advances in wave propagation algorithms, wavefield storage, and hardware acceleration are some of the main challenges concerning RTM \citep{zhou2018reverse}. For instance, the most effective non-reflecting boundary condition, Perfectly Matched Layer (PML), demands additional partial differential equations (PDEs) to be solved on artificial layers around the domain \citep{komatitsch2007unsplit, pasalic2010convolutional} to deal with unwanted reflections due to truncated domains. On the other hand, the forward-propagated wavefield concerning the RTM technique is a bottleneck due to the amount of information that has to be stored on a disk to build the imaging condition \citep{zhou2018reverse}. Besides, because RTM is time-consuming and data-intensive \citep{barbosaenhancing, qawasmeh2017performance, serpa2021energy},  RTM needs to be developed to take advantage of computer hardware technologies such as CPUs equipped with multi-processors, graphics processing units (GPUs), field-programmable gate arrays (FPGAs), and vector processors (VPs).

Efficient non-reflecting boundary conditions such as the PML can demand excessive numerical calculations beyond the wave equations \citep{li2020efficient}. A way to overcome this issue is to use the random boundary conditions (RBC) proposed by \citet{clapp2009reverse}. Thus, instead of suppressing unwanted waves by inserting new equations into the problem, the methodology proposed by \citet{clapp2009reverse} is based on exploring low correlations with non-coherent signals coming from an artificial boundary with random velocities. Strategies to diminish input/output (I/O) related to the forward-propagated wavefield storage or its reconstruction are presented by \citep{de1986migraccao, symes2007reverse, clapp2008reverse, clapp2009reverse, sun2013two, nguyen2015five, barbosaenhancing, li2020efficient}. Among them, \citet{sun2013two} presented two strategies to reduce data storage, where one is based on the Nyquist sampling theorem, and the second one uses a lossless compression algorithm. In this sense, \citet{barbosaenhancing} studied the numerical impact of applying lossless and lossy compression to the forward-propagated wavefield of the RTM. They show that the careful use of high levels of data compression can significantly reduce the storage demand without hampering the final seismic images. Instead of storing the wavefield, its reconstruction is a viable possibility. This can be done by checkpoint methods \citep{de1986migraccao, symes2007reverse}, using wavefield recording around the boundary \citep{clapp2008reverse, nguyen2015five}, or by initial value reconstruction (IVR) \citep{clapp2009reverse, nguyen2015five, li2020efficient}. \citet{nogueira20213d}, on the other hand, developed a dynamic approach for the RTM that delimits the computational domain by the seismic wavefront. According to the authors, the proposed strategy leads to memory savings and reduced processing times compared to the conventional implementation of the RTM. 

Independent of the RTM implementation strategy, all can use HPC techniques to boost their performance. Aiming to develop portable high-level directive-based codes across heterogeneous platforms for seismic imaging applications, \citet{qawasmeh2017performance} implemented the seismic modeling and RTM on a single as well multiple GPUs using a hybrid MPI+OpenACC approach. \citet{serpa2021energy} evaluated three different computational optimizations based on multicore and GPU architectures and investigated the performance, energy efficiency, and portability of the codes. Nevertheless, the storage demand issue remained in the RTM-based GPU implementations presented by the earlier research. For this, \citet{liu2012issues} implemented the RTM with RBC to diminish the storage demand need in migration algorithms showing that such a strategy is beneficial for GPU implementations. The GPU computational implementation with the RBC technique was coded in CUDA and tested only for 2-D RTM applications.

In this context, we developed a wave propagation modeler and an RTM approach for 2-D and 3-D environments that explore the main characteristics of the RBC to mitigate calculations on the artificial boundaries. Besides, our RTM implementation takes advantage of the RBC's non-dissipative energy in the system to reconstruct the full forward-propagated wavefield with minimum storage by the IVR technique. Our implementation is particularly suited for the new generation of vector processors, the NEC SX-Aurora TSUBASA. Computational times and disk storage results for two algorithmic choices (with and without IO) are compared in different computational platforms: a CPU cluster, a CPU-GPU cluster, and the vector processor. We show that our computational implementations are efficient, scalable, and portable with minimum interference on the optimized baseline code.

The remainder of the work is organized by introducing the mathematical background concerning the seismic modeling and RTM technique in sections \ref{sec.seis_modeling_math} and \ref{sec.rtm_math}. Section \ref{sec.comput_implem_optm} details the computational implementation for the seismic modeling and RTM along with optimizations on NEC SX-Aurora TSUBASA and NVIDIA Volta V100 platforms to highlight the differences between the two. In section \ref{sec.numerical_results}, we present numerical experiments where we expose the execution time requirements, speedups, and hard disk demand for each computational implementation, as well as the seismic modeling and RTM outcomes. The paper ends with a summary of our main findings in section \ref{sec.main_findings}.

\section{Seismic Modeling} \label{sec.seis_modeling_math}

In geophysical applications, seismic modeling is referred to as simulating the wave propagation in the Earth subsurface \citep{tago2012modelling}. Understanding the propagation of seismic waves is one of the cornerstones of geophysical data processing, such as RTM and FWI \citep{tago2012modelling, igel2017computational}. For an acoustic medium, the wave equation is described by the second-order partial differential equation as follows,

\begin{equation} \label{eq.secondOrder}
  \nabla^{2} p\left(\textbf{r},t \right)- \frac{1}{v^{2}\left(\textbf{r}\right)}\frac{\partial^{2} p\left(\textbf{r},t\right)}{\partial t^{2}} = f\left(\mathbf{r}_{s},t\right),
\end{equation}
where, $p$ is the pressure, $v$ the velocity for the compressional wave, $\textbf{r}$ the spatial coordinates, $t$ the time in $[0,T]$, and $f\left(\mathbf{r}_{s},t\right)$ the seismic source at the position $\textbf{r}_{s}$. The pressure $p$ is defined in a domain $\Omega \subset \mathbb{R}^{n_{sd}}$, $n_{sd} = 2, 3$. The second-order differential equation (\ref{eq.secondOrder}) needs initial and boundary conditions. A natural initial condition is to define $p\left(\textbf{r},0 \right) = \partial p\left(\textbf{r},0 \right) / \partial t = 0$ for $\textbf{r} \in \Omega$. Lastly, we set $p\left(\textbf{r},t \right) = 0$ on $\partial \Omega \in \mathbb{R}^{n_{sd}-1}$, where $\partial \Omega$ is the domain boundary.

\section{Reverse Time Migration} \label{sec.rtm_math}
Reverse Time Migration (RTM) is a depth migration technique based on the two-way wave equation, and an imaging condition \citep{zhou2018reverse}. Solving the wave equation twice to build the imaging condition is necessary. The first solution, called forward-propagated wavefield, can be obtained by solving the equation (\ref{eq.secondOrder}). The second solution is obtained by solving the following equation:

\begin{equation} \label{eq.secondOrderBackward}
\nabla^{2} \Bar{p}\left(\textbf{r},\tau\right) - \frac{1}{v^{2}\left(\textbf{r}\right)}\frac{\partial^{2} \Bar{p}\left(\textbf{r},\tau\right)}{\partial \tau^{2}} = s\left(\mathbf{r}_{r},\tau\right),
\end{equation}
where, $\Bar{p}$ is the backward-propagated wavefield, $s\left(\mathbf{r}_{r},\tau\right)$ is the seismogram recorded at the receivers positions $\mathbf{r}_{r}$, and $\tau = T - t$ is the reversal time evolution defined as in \citet{givoli2014time}, where $\tau \in [0, T]$. $\Bar{p}$ is also defined in $\Omega \subset \mathbb{R}^{n_{sd}}$, and corresponding initial, and boundary conditions should be set.

Once we have the forward- and backward-propagated wavefields, the imaging condition can be calculated as:

\begin{equation} \label{eq.imageCondition}
I\left(\textbf{r}\right) = \frac{\int_{0}^{T} p\left(\textbf{r},t\right) \, \Bar{p}\left(\textbf{r},\tau\right)\ dt}{\int_{0}^{T} \left[ p\left(\textbf{r},t\right)\right]^{2} dt},
\end{equation}
where $I\left(\textbf{r}\right)$ is called source-normalized cross-correlated imaging condition. The source-normalized cross-correlation image in equation (\ref{eq.imageCondition}) has the same unit, scaling, and sign of the reflection coefficient \citep{zhou2018reverse}.

\section{Computational Implementation and Optimizations} \label{sec.comput_implem_optm}

Our numerical implementation of seismic modeling and RTM employs the explicit Finite Difference Method (FDM) to solve the acoustic wave equation. The finite difference stencil for equations (\ref{eq.secondOrder}) and (\ref{eq.secondOrderBackward}) are 8th-order in space and 2nd-order in time. Thus, the numerical discretization leads to the discrete version of the velocity field, forward-propagated wavefield, backward-propagated wavefield, seismic source, and seismograms represented by the vectors $\mathbf{v}$, $\mathbf{p}$, $\Bar{\mathbf{p}}$, $\mathbf{f}$, and $\mathbf{s}$, respectively. For the 3-D case, the vectors $\mathbf{v}$, $\mathbf{p}$, $\Bar{\mathbf{p}}$ have the dimension $N = N_{x} \times N_{y} \times N_{z}$, where $N_{x}$, $N_{y}$ and $N_{z}$ are the number of grid points in each Cartesian direction. On the other hand, the seismogram is a vector of size $N_{rec} \times \left(N_{t} + 1 \right)$, where $N_{rec}$ is the number of receivers, and $N_{t} = T / \Delta t$, with $\Delta t$ the time step. Lastly, the seismic source $\mathbf{f}$ has dimension $N_{t}$ for each shot. The details of seismic modeling and RTM algorithms are presented in sections \ref{sec.impleSeisModel}, \ref{sec.impleRTM}, and \ref{sec.wavereconstruction}. Section \ref{sec.ImpleHPC} presents the computational optimization and parallelization developed for the SX-Aurora TSUBASA vector engine and, for comparison purposes, GPUs systems.

\subsection{Seismic Modeling} \label{sec.impleSeisModel}

The description of the seismic modeling algorithm is quite simple once we have the discretized version of the wave equation. Algorithm \ref{alg.seisModeling} shows that for the acoustic wave equation, only two inputs are needed. The first one is a velocity field $\mathbf{v}$ representing the spatial velocity distribution for the compressional wave (P-wave), and the second is a vector $\mathbf{f}$ containing information about the seismic signature, called seismic source, responsible for initiating the wave propagation through the medium.

\begin{algorithm}
    \caption{Seismic modeling}\label{alg.seisModeling}
        \begin{algorithmic}[1]
            \Require $\mathbf{v}$, and $\mathbf{f}$
            \Function{seismic\_modeling}{ vector $\mathbf{v}$, vector $\mathbf{f}$ }
                \State read $\mathbf{v}$, and $\mathbf{f}$
                \For{$shot\_id = 1$ to $N_{shots}$}
                    \State initialize $n_{t} = 0$
                    \State apply initial conditions for $i_{t} = 0$
                    \For{$i_{t} = 1$ to $N_{t}$}
                        \State $n_{t} = n_{t} + i_{t} * \Delta t$
                        \State solve equation \eqref{eq.secondOrder}
                        \State record seismogram signals near surface
                    \EndFor
                    \State store $\mathbf{s}_{shot\_id}$
                \EndFor
            \EndFunction
        \end{algorithmic}
\end{algorithm}

The wave equation propagation is solved over a temporal loop (the inner loop of Algorithm \ref{alg.seisModeling}) for each $shot\_{id}$ (loop in line 3). The shot refers to the seismic source that starts the wave propagation, and each one is localized in the domain represented by the finite-difference grid. The algorithm finishes recording a seismogram $\mathbf{s}_{shot\_id}$ associated with each shot. A computational implementation of absorbing boundary conditions (ABCs) leads to spurious reflections on the truncated domain. Among the several options in the literature, the Convolutional Perfectly Matched Layer (CPML) \citep{komatitsch2007unsplit, pasalic2010convolutional} and the damping factors for plane waves introduced by \citet{cerjan1985nonreflecting} are the most common. Although unusual in wave propagation simulation studies, the RBCs, first introduced by \citet{clapp2009reverse}, can also be employed in seismic imaging methods based on the two-way wave equation, such as the RTM and FWI \citep{clapp2008reverse, clapp2009reverse, nguyen2015five, li2020efficient}. Further discussions about the use of the RBC with the RTM are made in the wavefield reconstruction section \ref{sec.wavereconstruction}.

\subsection{Reverse Time Migration} \label{sec.impleRTM}

Algorithm \ref{alg.seisModeling} detailed in section \ref{sec.impleSeisModel} is the kernel for the RTM algorithm presented in Algorithm \ref{alg.rtmuq}. Again, the two inputs are the velocity field and the seismic source. Besides, the RTM needs a set of seismograms, $\{ \mathbf{s}_{1}, \cdot \cdot \cdot, \mathbf{s}_{N_{shots}}\}$ that contains information about the medium reflectivity. The computation of the imaging condition uses the forward-propagated, and backward-propagated wavefield solutions to build the migrated seismic section that stacks the partial results over time $\left( \mathbf{I}_{\sum n_{\tau}} \right)$, and over the number of seismograms $\left( \mathbf{I}_{\sum shot\_id} \right)$. We compute the forward-propagated wavefield by solving the wave equation with the independent term being the seismic source and storing it in disk for further access (step 10 in red). On the other hand, the recorded seismograms induce the computation of the backward-propagated wavefield. At the end of Algorithm \ref{alg.rtmuq}, we obtain the discrete seismic image $I \in \mathbb{R}^{N_{x} \times N_{y} \times N_{z}}$, where the amplitude variations represent physical properties changes.

\begin{algorithm}
    \caption{Reverse Time Migration}\label{alg.rtmuq}
        \begin{algorithmic}[1]
            \Require $\mathbf{v}$, $\{ \mathbf{s}_{1}, \cdot \cdot \cdot, \mathbf{s}_{N_{shots}}\}$, and $\mathbf{f}$
            \Function{rtm}{ vector $\mathbf{v}$, vectors $\{ \mathbf{s}_{1}, \cdot \cdot \cdot, \mathbf{s}_{N_{shots}}\}$, vector $\mathbf{f}$ }
                \State read $\mathbf{v}$, $\mathbf{f}$, and $\{ \mathbf{s}_{1}, \cdot \cdot \cdot, \mathbf{s}_{N_{shots}}\}$
                \State initialize image condition $\mathbf{I}_{\sum shot\_id} = 0$
                \For{$shot\_id = 1$ to $N_{shots}$}
                    \State initialize $n_{t} = 0$
                    \State apply initial conditions for $i_{t} = 0$
                    \For{$i_{t} = 1$ to $N_{t}$}
                        \State $n_{t} = n_{t} + i_{t} * \Delta t$
                        \State solve equation \eqref{eq.secondOrder} \Comment{source wavefield}
                        \State \textcolor{red}{store $\mathbf{p}_{n_{t}}$ for all $n_{t}$}
%                       \quad \forall t \in 0 \leq t \leq T$
                    \EndFor
                    \State initialize $n_{\tau} = 0$, and $\mathbf{I}_{\sum \tau} = 0$
                    \State apply initial conditions for $i_{\tau} = 0$
                    \For{$i_{\tau} = 1$ to $N_{t}$}
                        \State $n_{\tau} = N_{t} - (n_{\tau} + i_{\tau} * \Delta \tau)$ \Comment{reverse time}
                        \State read $\mathbf{p}_{n_{\tau}}$, and $\mathbf{s}_{shot\_id}$
                        \State solve equation \eqref{eq.secondOrderBackward} \Comment{receiver wavefield}
                        \State calculate $\mathbf{I}_{\sum n_{\tau}} = \mathbf{I}_{\sum n_{\tau}} + \left( \mathbf{p}_{n_{\tau}} \Bar{\mathbf{p}}_{n_{\tau}} \right) / \left( \mathbf{p}_{n_{\tau}}\mathbf{p}_{n_{\tau}} \right)$ \Comment{imaging condition}
                    \EndFor
                    \State stack $\mathbf{I}_{\sum shot\_id} = \mathbf{I}_{\sum shot\_id} + \mathbf{I}_{\sum n_{\tau}}$ \Comment{stacking}
            \EndFor
            \State $\mathbf{I} \leftarrow \mathbf{I}_{\sum shot\_id}$
            \State store $\mathbf{I}$
            \EndFunction
        \end{algorithmic}
\end{algorithm}

The RTM implementation presented in Algorithm \ref{alg.rtmuq} is one of the simplest ways to build the cross-correlated imaging condition. The algorithm involves calculating the wave equation twice and storing the forward-propagated wavefield to access it in the reverse way to correlate with the backward-propagated wavefield. However, storing and accessing the forward-propagated wavefield is computationally demanding. In this work, we implement the wavefield reconstruction based on the IVR technique with pseudorandomized wavefield proposed by \citet{clapp2009reverse} to deal with persistent storage of the forward-propagated wavefield. The next section \ref{sec.wavereconstruction} details the IVR strategy and modifications for Algorithm \ref{alg.rtmuq}.

\subsection{Wavefield Reconstruction} \label{sec.wavereconstruction}

The basic RTM implementation as presented in Algorithm \ref{alg.rtmuq} suffers from persistent I/O due to the need to store the forward-propagated wavefield in disk for further access to calculate the imaging condition. One way to overcome this issue, explored in this work, is to reconstruct the forward-propagated wavefield from information generated during the first part of the RTM, that is, forward wave propagation \citep{nguyen2015five}. To reconstruct the forward-propagated wavefield, we implement the IVR methodology first explored by \citet{de1986migraccao} and \citet{symes2007reverse}. The IVR proposed by \citet{symes2007reverse} stores temporary states of the wavefield known as checkpoints. Such states are after used for recursive recomputations of the forward-propagated wavefield. On the other hand, \citet{de1986migraccao} uses a single checkpoint to initiate the backpropagation of the wavefield. However, using this concept with non-reflective boundary conditions can result in an inefficient reconstruction of the forward-propagated wavefield due to signal attenuation in the boundary \citep{silva2012modelagem}. The complete reconstruction of the wavefield can be achieved by keeping all energy in the system. However, unwanted signals come from the boundary due to the absence of attenuated layers on the boundaries used to simulate truncated domains. One way to overcome this issue is generating incoherent signals coming from the boundary as explored in \citet{clapp2009reverse} by introducing boundaries with randomized velocities.

The Random Boundary Condition (RBC) proposed by \citet{clapp2009reverse} is based on the idea that what matters for the calculation of the RTM imaging condition is the coherent reflections coming from the boundaries. Thus, \citet{clapp2009reverse} proposed to introduce a random component to the velocity field at the boundaries. Notice that the random velocity field has to respect the numerical stability constraint of the FDM. It is expected that the random forward-propagated wavefield coming from the boundaries does not coherently correlate with the backward-propagated wavefield. Besides, a smoother transition from the inner domain to the boundaries is ideal. The smooth transition will avoid unwanted immediate reflections of the randomized area. One way to build a smooth transition area is by multiplying coefficients $c_{i}$ to the random vector velocity $\mathbf{v}$ in the normal direction to the boundaries, where the index $i \in [1, \cdot \cdot \cdot, N_{a}]$ with $N_{a}$ been the size thickness of the boundaries. The coefficients are responsible for slowing down the velocities values, and \citet{silva2012modelagem} showed that values between the linear and Gaussian functions, represented by equations (\ref{eq.factors_rbc_linear}) and (\ref{eq.factors_rbc_gaussian}), build the best coefficients, that is,
\begin{equation} \label{eq.factors_rbc_linear}
g(x) = (N_{a} - x) \left( \frac{1}{N_{a} - 1} \right),
\end{equation}

\begin{equation} \label{eq.factors_rbc_gaussian}
h(x) = \exp{\left( -120 (x - 1)^{2} \left( \frac{1}{N_{a} - 1} \right)^{2} \right)},
\end{equation}
where, $x \in [1, N_{a}]$. Let \textbf{g}, and \textbf{h} be the discrete version of the functions $g(x)$, and $h(x)$ after numerical discretization, thus the damping coefficients assume values in $ \textbf{h} \leq c_{i} \leq \textbf{g}$ for $i \in [1, \cdot \cdot \cdot, N_{a}]$.

It is worth mentioning that this strategy is effective and does not impose an extra cost on the wave equation calculation. An alternative way to avoid coherent signals coming from the boundaries is presented by \citet{li2020efficient}, where they used an extra viscoacoustic wave equation in the boundaries to attenuate the wavefield. In this work, we employ the strategy presented by \citet{silva2012modelagem}. Details of the RBC algorithm can be observed in \citet{clapp2009reverse}. Here, we will describe the modifications for Algorithm \ref{alg.rtmuq} aiming to eliminate the storage requirements of the forward-propagated wavefield.

First, we need a third second-order wave equation as follows:

\begin{equation} \label{eq.secondOrderForwardReconstruction}
\nabla^{2} p^{R}\left(\textbf{r},\tau\right) - \frac{1}{v^{2}\left(\textbf{r}\right)}\frac{\partial^{2} p^{R}\left(\textbf{r},\tau\right)}{\partial \tau^{2}} = 0,
\end{equation}
where $p^{R}$ is the reconstructed forward-propagated wavefield defined in $\Omega \subset \mathbb{R}^{n_{sd}}$. Boundary conditions can be set as equations (\ref{eq.secondOrder}), and (\ref{eq.secondOrderBackward}), that is $p^{R}\left(\textbf{r},t \right) = 0$ on $\partial \Omega$. Lastly, the initial conditions are set as $p^{R}\left(\textbf{r},0 \right) = p\left(\textbf{r},T \right)$, and $\partial p^{R}\left(\textbf{r},0 \right) / \partial t = \partial p\left(\textbf{r},T \right) / \partial t$ after solving equation \ref{eq.secondOrder}, and $\tau = T - t$ is the reversal time.

Algorithm \ref{alg.rtmwr} highlights in blue the main modifications in the basic RTM algorithm. We use the vector $\mathbf{p}^{R}$ to represent the finite difference discretization of equation (\ref{eq.secondOrderForwardReconstruction}). The first part of the RTM with wavefield reconstruction calculates the forward-propagated wavefield, and the last two moments of the wavefield are stored (line 11). After reading the stored wavefield moments, the second part of the algorithm that calculates the backward-propagated wavefield also calculates the reconstruction of the forward-propagated wavefield $\mathbf{p}^{R}$ by solving equation (\ref{eq.secondOrderForwardReconstruction}). Thus, the modified algorithm stores only two panels of forward-propagated wavefield instead of all panels for each $n_{t}$. This strategy comes with the additional cost of solving one extra wave equation.

\begin{figure}[ht]
  \centering
  \includegraphics[scale=.35]{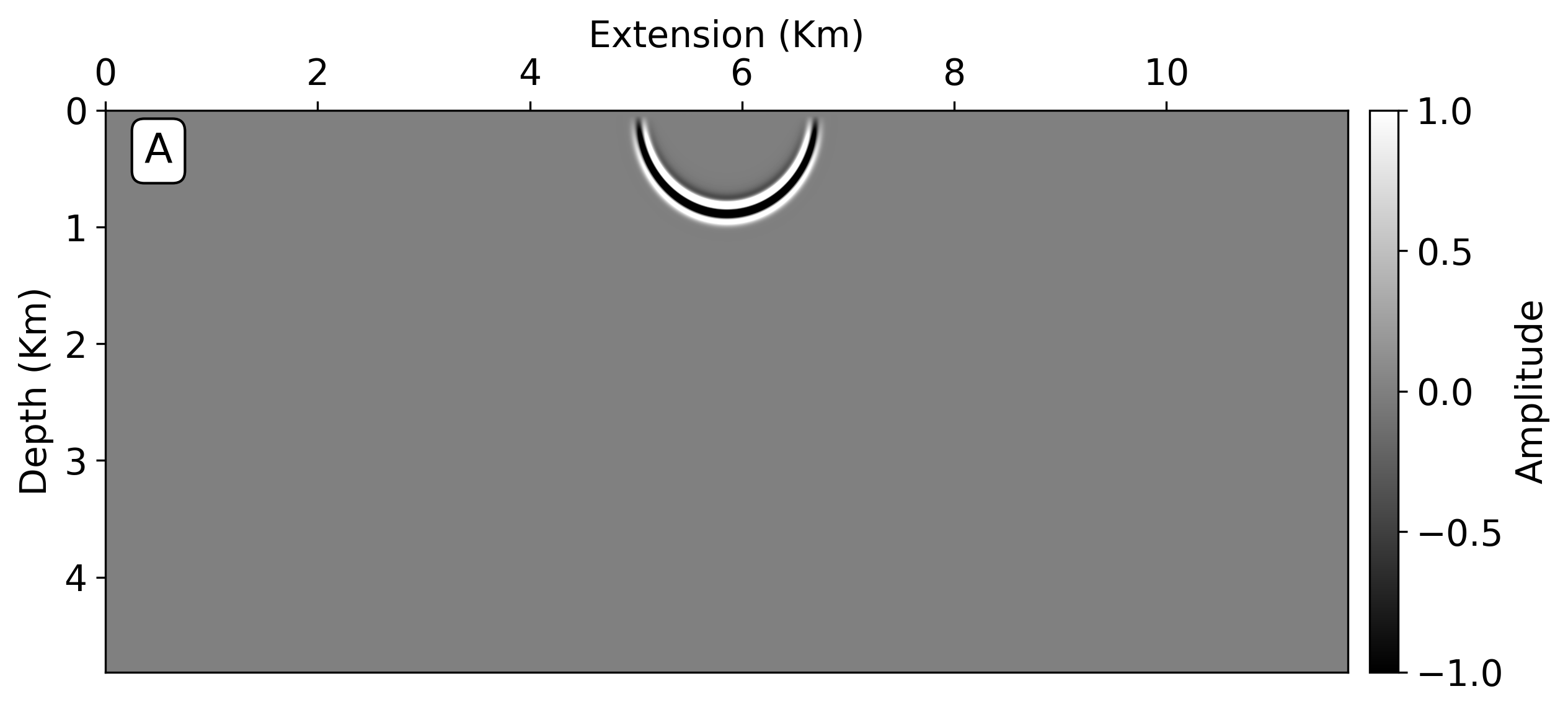}
  \includegraphics[scale=.35]{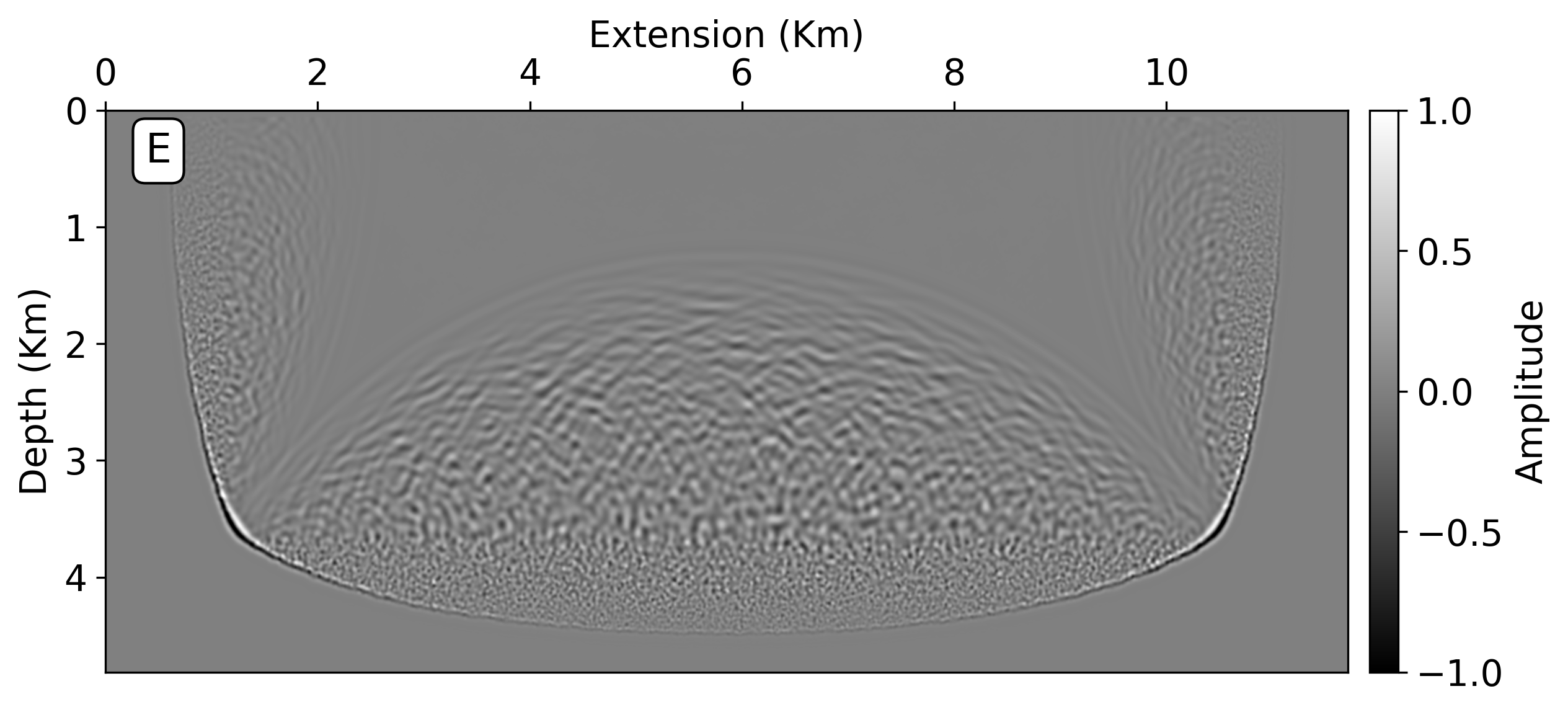}
  \includegraphics[scale=.35]{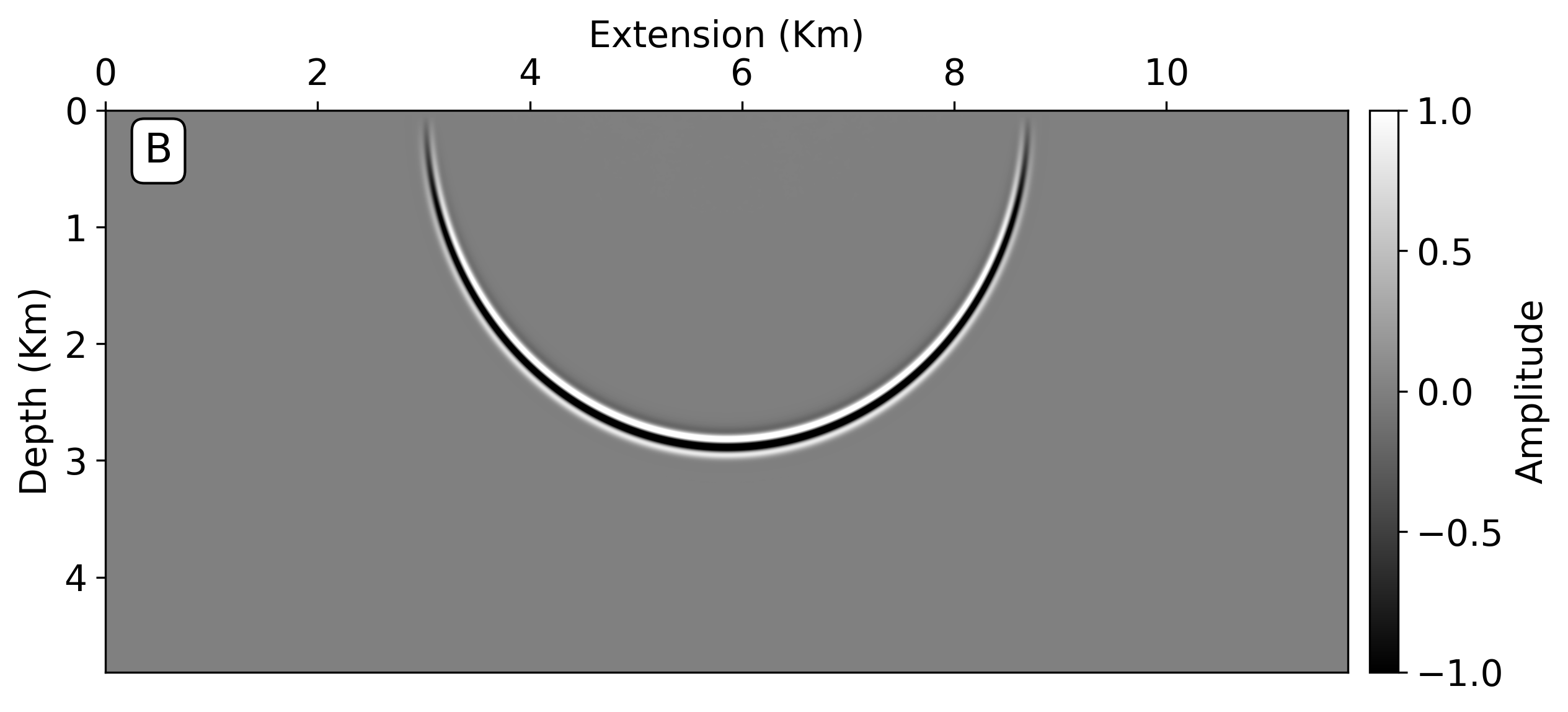}
  \includegraphics[scale=.35]{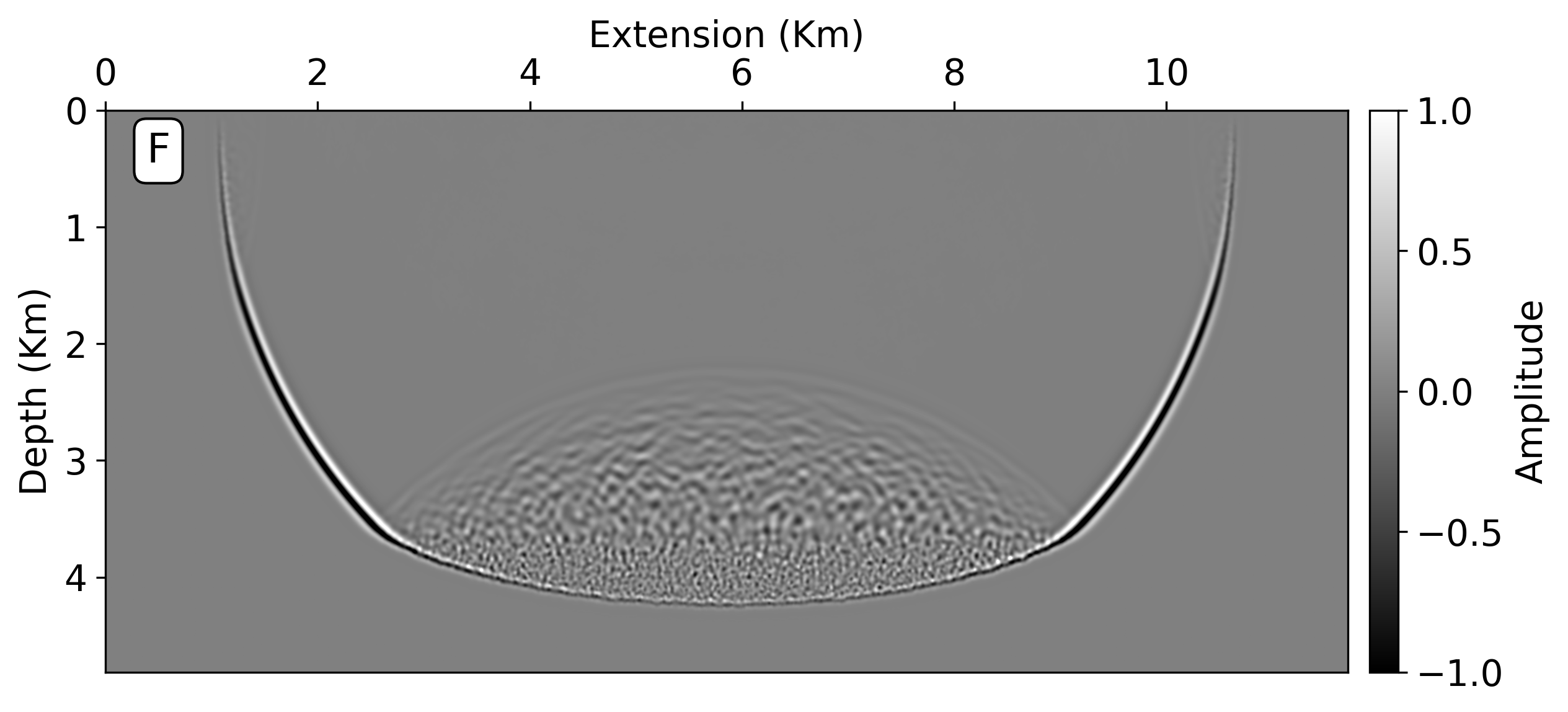}
  \includegraphics[scale=.35]{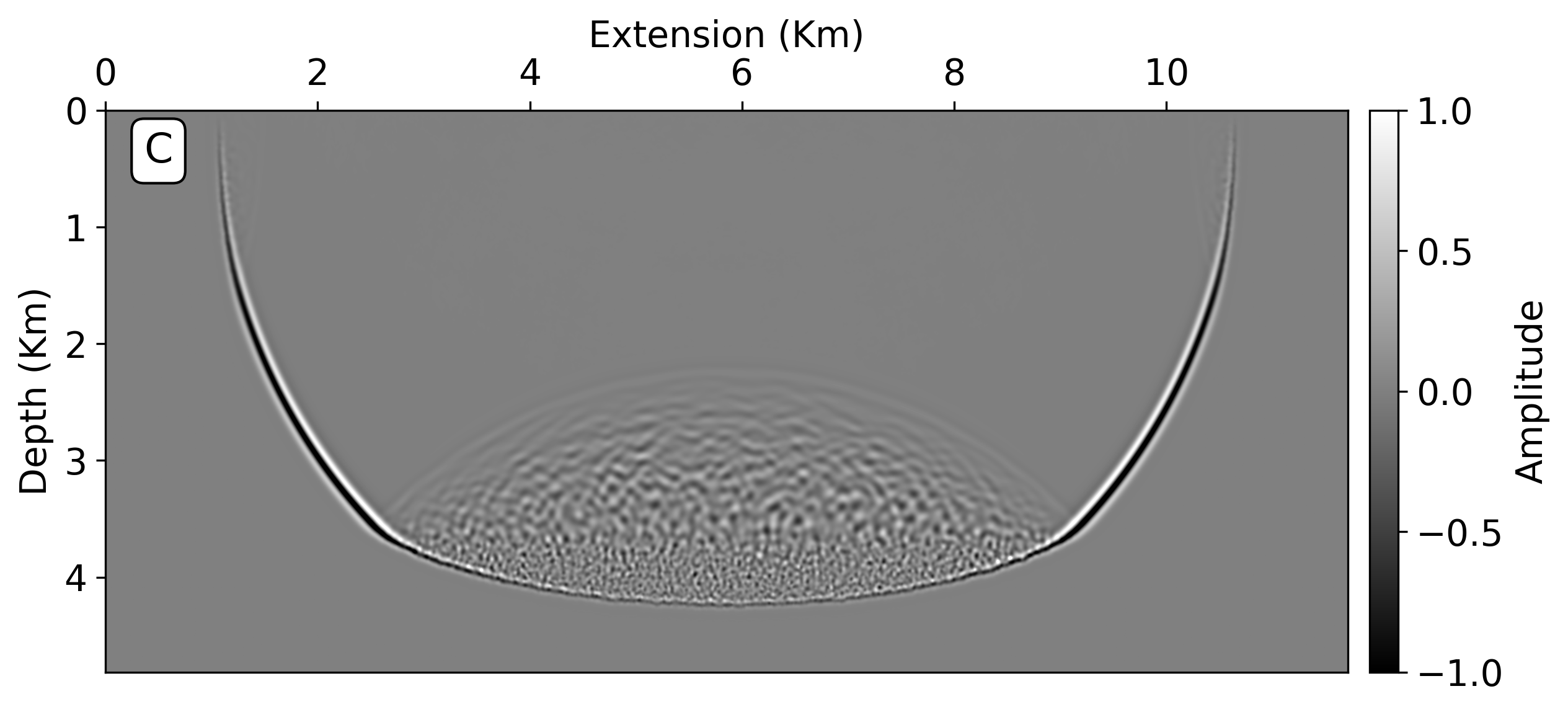}
  \includegraphics[scale=.35]{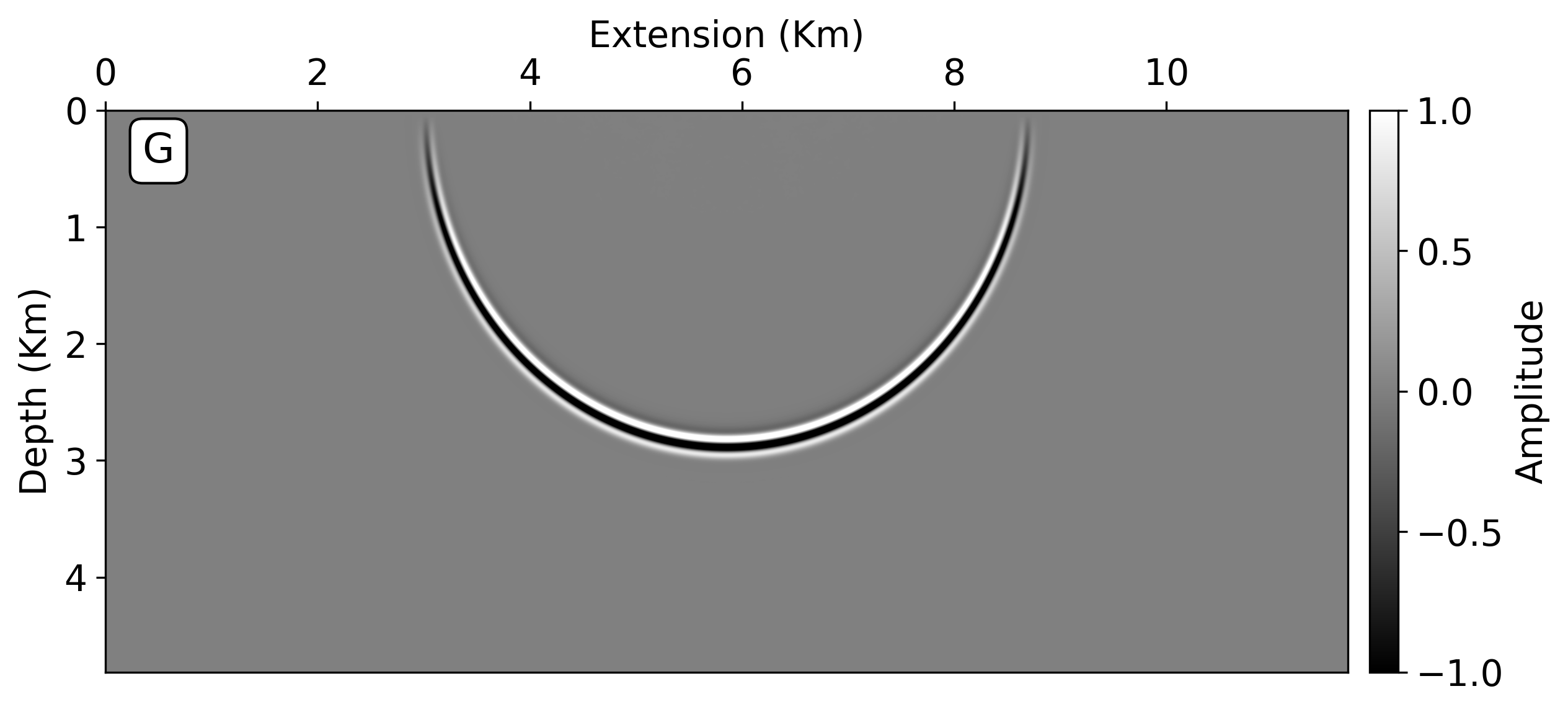}
  \includegraphics[scale=.35]{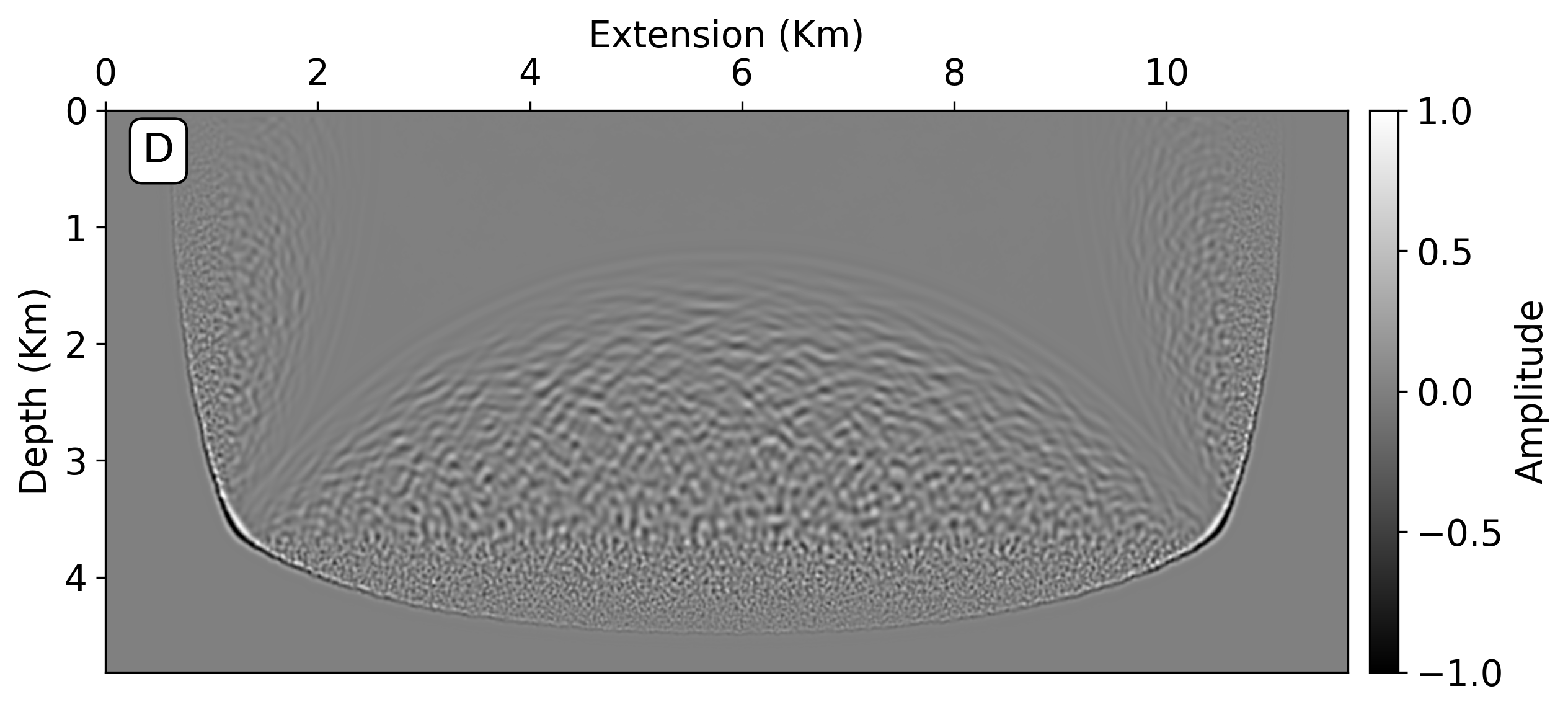}
  \includegraphics[scale=.35]{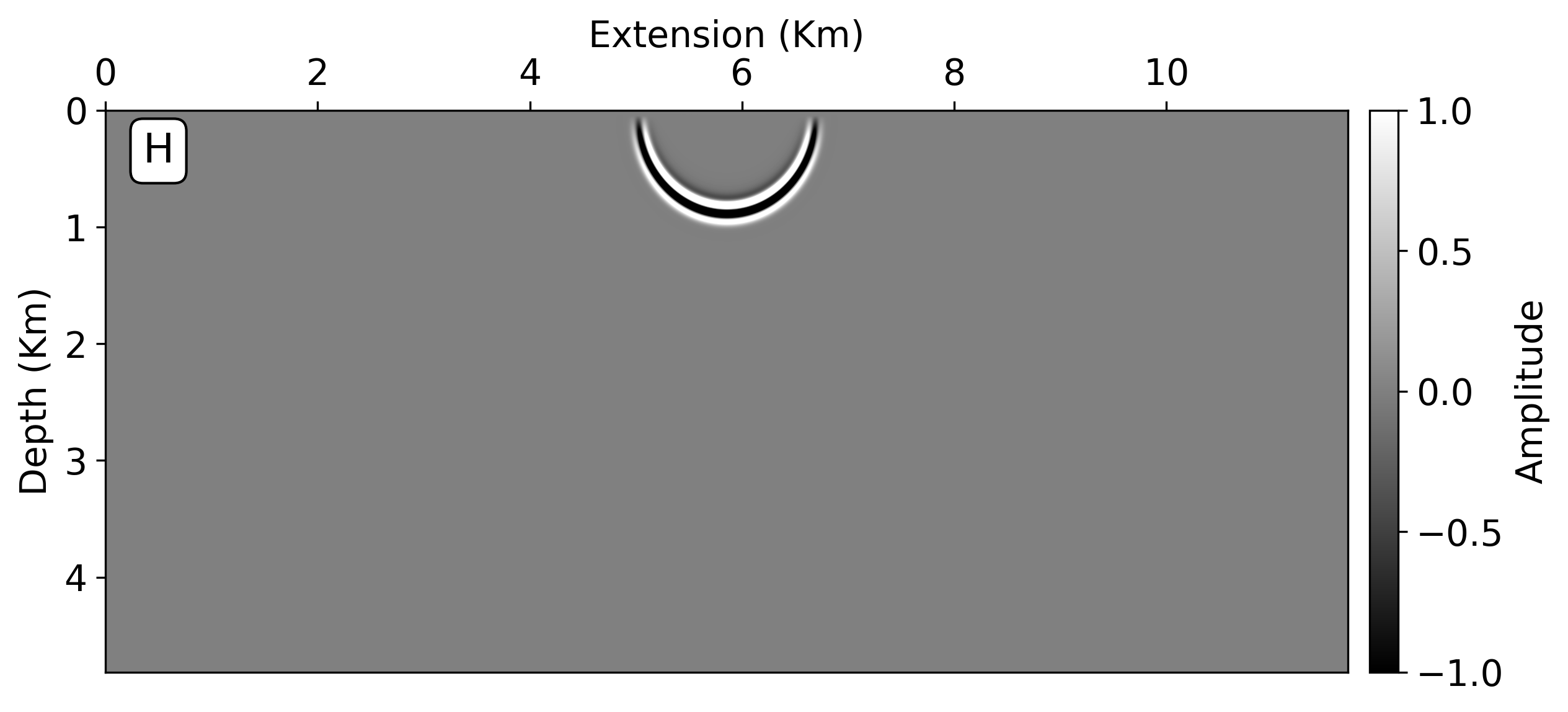}
  \caption{The left column shows the representation of the forward-propagated wavefield for the instants 0.25s (A), 1.5s (B), 2.5s (C), and 3.0s (D). The right column shows the representation of the wavefield reconstruction of the forward-propagated wavefield for instants  0.25s (E), 1.5s (F), 2.5s (G), and 3.0s (H). The results were provided by Algorithm \ref{alg.rtmwr}.}
  \label{fig:propagation_reconstruction}
\end{figure}

Figure \ref{fig:propagation_reconstruction} shows the representation of the propagation of the forward wavefield (Figures 1(A), 1(B), 1(C), and 1(D)) and its reconstruction (Figures 1(E), 1(F), 1(G), and 1(H)) in a constant velocity field of 2000 m/s based on Algorithm \ref{alg.rtmwr}. The experiment from Figure \ref{fig:propagation_reconstruction} implements the RBC, and, thus, it is possible to see the incoherent signals coming from the boundaries. Besides, the computational implementation for the Algorithm \ref{alg.rtmwr} maintains all energy inside the domain, allowing the complete reconstruction of the forward-propagated wavefield.

\begin{algorithm}
    \caption{Reverse Time Migration with Wavefield Reconstruction}\label{alg.rtmwr}
        \begin{algorithmic}[1]
            \Require $\mathbf{v}$, $\{ \mathbf{s}_{1}, \cdot \cdot \cdot, \mathbf{s}_{N_{shots}}\}$, and $\mathbf{f}$
            \Function{rtm}{ vector $\mathbf{v}$, vectors $\{ \mathbf{s}_{1}, \cdot \cdot \cdot, \mathbf{s}_{N_{shots}}\}$, vector $\mathbf{f}$ }
                \State read $\mathbf{v}$, $\mathbf{f}$, and $\{ \mathbf{s}_{1}, \cdot \cdot \cdot, \mathbf{s}_{N_{shots}}\}$
                \State \textcolor{blue}{create a RBC as Algorithm 2 from \citet{clapp2009reverse}}
                \State initialize image condition $\mathbf{I}_{\sum shot\_id} = 0$
                \For{$shot\_id = 1$ to $N_{shots}$}
                    \State initialize $n_{t} = 0$
                    \State apply initial conditions for $i_{t} = 0$
                    \For{$i_{t} = 1$ to $N_{t}$}
                        \State $n_{t} = n_{t} + i_{t} * \Delta t$
                        \State solve equation \eqref{eq.secondOrder} \Comment{source wavefield}
                        \State \textcolor{blue}{store $\mathbf{p}_{n_{t}}$ for $N_{t-1}$, and $N_{t}$}
%                       \quad \forall t \in 0 \leq t \leq T$
                    \EndFor
                    \State initialize $n_{\tau} = 0$, and $\mathbf{I}_{\sum \tau} = 0$
                    \State \textcolor{blue}{read $\mathbf{p}_{N_{t}}$, and $\mathbf{p}_{N_{t-1}}$}
                    \State apply initial conditions for $i_{\tau} = 0$
                    \For{$i_{\tau} = 1$ to $N_{t}$}
                        \State $n_{\tau} = N_{t} - (n_{\tau} + i_{\tau} * \Delta \tau)$ \Comment{reverse time}
                        \State read $\mathbf{s}_{shot\_id}$
                        \State solve equation \eqref{eq.secondOrderBackward} \Comment{receiver wavefield}
                        \State \textcolor{blue}{solve equation \eqref{eq.secondOrderForwardReconstruction}} \Comment{\textcolor{blue}{wavefield reconstruction}}
                        \State calculate $\mathbf{I}_{\sum n_{\tau}} = \mathbf{I}_{\sum n_{\tau}} + \left( \mathbf{p}^{R}_{n_{\tau}} \Bar{\mathbf{p}}_{n_{\tau}} \right) / \left( \mathbf{p}^{R}_{n_{\tau}}\mathbf{p}^{R}_{n_{\tau}} \right)$ \Comment{imaging condition}
                    \EndFor
                    \State stack $\mathbf{I}_{\sum shot\_id} = \mathbf{I}_{\sum shot\_id} + \mathbf{I}_{\sum n_{\tau}}$ \Comment{stacking}
            \EndFor
            \State $\mathbf{I} \leftarrow \mathbf{I}_{\sum shot\_id}$
            \State store $\mathbf{I}$
            \EndFunction
        \end{algorithmic}
\end{algorithm}

\subsection{Vector Processor and OpenACC Implementations} \label{sec.ImpleHPC}

We develop versions for the seismic modeling and RTM on a traditional scalar CPU to get an optimized baseline implementation. Our optimized baseline implementations take advantage of the Single-Instruction-Multiple-Data (SIMD) model and memory alignment allocation to ensure vectorization \citep{barbosa2020workflow}. Both versions were implemented in C language, and they were used as the start point for the vector processor and GPU implementations, based on OpenACC directives. The sections \ref{sec.vectorImple}, and \ref{sec.openACCImple} present the main vector processor and GPU characteristics and details the computational optimizations and parallelization for the seismic applications of this work.

\subsubsection{Vector Processor Implementation} \label{sec.vectorImple}

Our computational vector processor implementation is directed toward the NEC SX-Aurora TSUBASA vector processor. The SX-Aurora TSUBASA architecture consists of a vector engine (VE) equipped with a vector processor and a vector host (VH). In this architecture, the VE runs the entire application, while the VH is responsible for processing system calls invoked by the application \citep{komatsu2018performance}. Besides, the architecture avoids frequent data transfers between the VE and its VH. Developing scientific applications for the SX-Aurora TSUBASA is straightforward because no special coding is required, and the developers do not need to take care of the system calls.

In this sense, the implementations for the Algorithms \ref{alg.seisModeling}, \ref{alg.rtmuq}, and \ref{alg.rtmwr} presented in sections \ref{sec.impleSeisModel}, \ref{sec.impleRTM}, and \ref{sec.wavereconstruction} do not have any special code modification with respect to the optimized serial version that is our baseline implementation. Although the NEC SX-Aurora TSUBASA provides support for OpenMP implementation, we have been using automatic parallelization and vectorization activated by the NEC compilations flags -O4, mparallel, -fivdep, and -mparallel-innerloop.

\subsubsection{OpenACC Implementation} \label{sec.openACCImple}

The GPU programming model, based on OpenACC directives, aims to provide an easier way for scientific applications coding \citep{qawasmeh2017performance, kushida2019acceleration}. Besides, compared to CUDA and OpenCL, OpenACC programming demands less coding efforts in heterogeneous environments with CPU+GPU \citep{qawasmeh2017performance, serpa2021energy}. The OpenACC implementation needs to deal with three main issues: CPU (host) calculations, GPU calculations, and communications to and from the GPU. Thus, any computational implementation must maximize the GPU computations and prevent communications between the host and GPU. 

Algorithm \ref{alg.seismicModelingGPU} details the host and GPU calculations and the communication between them for the seismic modeling. The first operations made by the host are data allocation followed by disk reading and storage of the velocity field and seismic source information in the vectors $\mathbf{v}$, and $\mathbf{f}$. These steps are shown in lines 2 and 3 in Algorithm \ref{alg.seismicModelingGPU}. Following, the main data, such as the vectors $\mathbf{v}$ and $\mathbf{f}$, are moved to the GPU (line 4). Lines 6 and 7 show the GPU operations for the wave equation calculation once the necessary information is transferred and allocated. The seismogram is the outcome of the wave equation simulation, and it is transferred from GPU to the host in line 9. The final operations of the host are seismogram storage and data deallocation in lines 10 and 11. Notice that for seismic modeling, only three communications are necessary. Because the velocity field and seismic source are information provided for the seismic modeling, two transfers are made from the host to GPU. In the end, the host stores the seismogram after its transfer from GPU to host. We use the ACC DATA COPYIN directive for transferring the data from the host to GPU. ACC DATA COPYOUT directive transfers the data from GPU to host. ACC DATA CREATE allocates necessary vectors in the GPU. For parallelization, we use the ACC LOOP directive.
\begin{algorithm}
    \caption{Seismic Modeling GPU Implementation}\label{alg.seismicModelingGPU}
        \begin{algorithmic}[1]
            \Require $\mathbf{v}$, and $\mathbf{f}$
            \Function{seismic\_modeling\_gpu}{ vector $\mathbf{v}$, vector $\mathbf{f}$ }
                \State \textcolor{green}{allocate data variables}                \Comment{\textcolor{green}{Host computations}}
                \State \textcolor{green}{read $\mathbf{v}$, and $\mathbf{f}$}
                \State \textcolor{blue}{Move data to GPU}               \Comment{\textcolor{blue}{Data transfer and allocation}}
                \For{time = first to last}
                    \State \textcolor{red}{solve wave equation (\ref{eq.secondOrder})}      \Comment{\textcolor{red}{GPU computations}}
                    \State \textcolor{red}{record seismogram}
                \EndFor
                \State \textcolor{blue}{update host with seismogram} \Comment{\textcolor{blue}{Data transfer and deallocation}}
                \State \textcolor{green}{store seismogram}      \Comment{\textcolor{green}{Host computations}}
                \State \textcolor{green}{deallocate data variables}
            \EndFunction
        \end{algorithmic}
\end{algorithm}

Most of the OpenACC implementations presented in Algorithm \ref{alg.seismicModelingGPU} for the seismic modeling are the same for the RTM algorithm. The differences between the seismic modeling and RTM algorithms are mainly related to data transfer. For instance, Algorithm \ref{alg.rtmAlg2GPU} details the OpenACC implementation for the RTM which implements the wavefield storage (Algorithm \ref{alg.rtmuq}). Again, we use three different colors to represent the host computations (green), data transfer (blue), and GPU calculations (red). The first part of Algorithm \ref{alg.rtmAlg2GPU} moves the source wavefield during its calculation from the GPU to the host and stores it in disk (lines 6 to 9). In general, storing the source wavefield in a disk is needed because the GPU memory or RAM is insufficient to store it. The second part of the RTM algorithm (lines 13 to 19) moves back the source wavefield from host to GPU, calculates the receiver wavefield, and builds the imaging condition (lines 15 to 18). Algorithm \ref{alg.rtmAlg2GPU} requires two data transfers for the velocity field and seismic source, $N_{shots}$ data transferring for the seismograms, and $2 \times N_{t}$ data transferring for the source wavefield.
\begin{algorithm}
    \caption{RTM GPU Implementation based on Algorithm \ref{alg.rtmuq}}\label{alg.rtmAlg2GPU}
        \begin{algorithmic}[1]
            \Require $\mathbf{v}$, $\mathbf{f}$, and $\{ \mathbf{s}_{1}, \cdot \cdot \cdot, \mathbf{s}_{N_{shots}}\}$
            \Function{rtm\_gpu}{ vector $\mathbf{v}$, vectors $\{ \mathbf{s}_{1}, \cdot \cdot \cdot, \mathbf{s}_{N_{shots}}\}$, vector $\mathbf{f}$ }
                \State \textcolor{green}{allocate data variables}                \Comment{\textcolor{green}{Host computations}}
                \State \textcolor{green}{read $\mathbf{v}$, and $\mathbf{f}$}
                \State \textcolor{blue}{Move data to GPU}               \Comment{\textcolor{blue}{Data transfer and allocation}}
                \For{time = first to last}
                    \State \textcolor{red}{solve wave equation \eqref{eq.secondOrder}}          \Comment{\textcolor{red}{GPU computations}}
                    \State \textcolor{red}{record source wavefield for every time}
                    \State \textcolor{blue}{move source wavefield to host} \Comment{\textcolor{blue}{Data transfer}}
                \State \textcolor{green}{store source wavefield for each time}      \Comment{\textcolor{green}{Host computations}}
                \EndFor
                \State \textcolor{green}{read seismograms}
                \State \textcolor{blue}{Move seismogram to GPU}         \Comment{\textcolor{blue}{Data transfer and allocation}}
                \For{time = first to last}
                    \State \textcolor{red}{set reversal time evolution}
                    \State \textcolor{green}{read source wavefield for each reversal time}       \Comment{\textcolor{green}{Host computations}}
                    \State \textcolor{blue}{Move source wavefield to GPU}               \Comment{\textcolor{blue}{Data transfer and allocation}}
                    \State \textcolor{red}{solve wave equation \eqref{eq.secondOrderBackward}}          \Comment{\textcolor{red}{GPU computations}}
                    \State \textcolor{red}{calculate imaging condition}
                \EndFor
                \State \textcolor{blue}{update host with seismic image}               \Comment{\textcolor{blue}{Data transfer and deallocation}}
                \State \textcolor{green}{store seismic image}      \Comment{\textcolor{green}{Host computations}}
                \State \textcolor{green}{deallocate all the data variables}
            \EndFunction
        \end{algorithmic}
\end{algorithm}

The OpenACC implementation based on Algorithm \ref{alg.rtmwr} is shown in Algorithm \ref{alg.rtmAlg3GPU}. Remember that Algorithm \ref{alg.rtmwr} implements the wavefield reconstruction, and one extra wave equation is required for that. Because of that, its computational implementation does not fully store the source wavefield, only the last two time-frames. The data transfer based on the OpenACC implementation occurs between the two main stages of the RTM technique and not during the temporal loops as the Algorithm \ref{alg.rtmAlg2GPU}. Thus, Algorithm \ref{alg.rtmAlg3GPU} requires only four data transfers between the GPU and host for the source wavefield. We use for both Algorithms \ref{alg.rtmAlg2GPU} and \ref{alg.rtmAlg3GPU} the same pragma directives that we use in seismic modeling. The ACC DATA COPYIN, ACC DATA COPYOUT for data transfer, ACC DATA CREATE for data allocation, and ACC LOOP for parallelization.
\begin{algorithm}
    \caption{RTM GPU Implementation based on Algorithm \ref{alg.rtmwr}}\label{alg.rtmAlg3GPU}
        \begin{algorithmic}[1]
            \Require $\mathbf{v}$, $\mathbf{f}$, and $\{ \mathbf{s}_{1}, \cdot \cdot \cdot, \mathbf{s}_{N_{shots}}\}$
            \Function{rtm\_gpu}{ vector $\mathbf{v}$, vectors $\{ \mathbf{s}_{1}, \cdot \cdot \cdot, \mathbf{s}_{N_{shots}}\}$, vector $\mathbf{f}$ }
                \State \textcolor{green}{allocate data variables}                \Comment{\textcolor{green}{Host computations}}
                \State \textcolor{green}{read $\mathbf{v}$, and $\mathbf{f}$}
                \State \textcolor{blue}{Move data to GPU}               \Comment{\textcolor{blue}{Data transfer and allocation}}
                \For{time = first to last}
                    \State \textcolor{red}{solve wave equation \eqref{eq.secondOrder}}          \Comment{\textcolor{red}{GPU computations}}
                    \State \textcolor{red}{record source wavefield for every time}
                \EndFor
                \State \textcolor{blue}{update host with the last two wavefield timeframes} \Comment{\textcolor{blue}{Data transfer and deallocation}}
                \State \textcolor{green}{read seismograms}
                \State \textcolor{blue}{Move seismogram and wavefield timeframes to GPU}        \Comment{\textcolor{blue}{Data transfer and allocation}}
                \For{time = first to last}
                    \State \textcolor{red}{set reversal time evolution}
                    \State \textcolor{red}{solve wave equation \eqref{eq.secondOrderBackward}}          \Comment{\textcolor{red}{GPU computations}}
                    \State \textcolor{red}{solve wave equation \eqref{eq.secondOrderForwardReconstruction}}
                    \State \textcolor{red}{calculate imaging condition}
                \EndFor
                \State \textcolor{blue}{update host with seismic image}               \Comment{\textcolor{blue}{Data transfer and deallocation}}
                \State \textcolor{green}{store seismic image}      \Comment{\textcolor{green}{Host computations}}
                \State \textcolor{green}{deallocate all the data variables}
            \EndFunction
        \end{algorithmic}
\end{algorithm}

\section{Numerical Results} \label{sec.numerical_results}
\begin{figure}[ht]
  \centering
  \includegraphics[width=\linewidth]{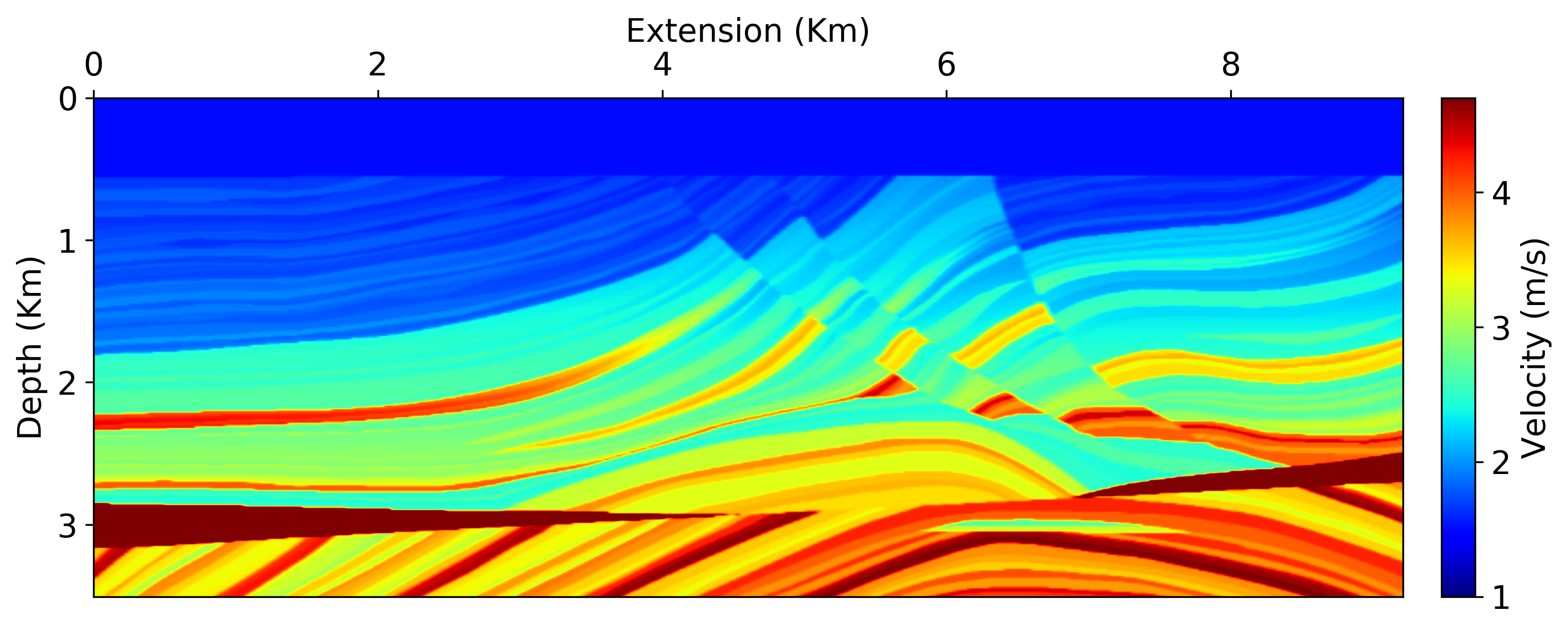}
  \caption{2-D Marmousi velocity model benchmark.}
  \label{fig:marmousi_bench}
\end{figure}

\begin{figure}[ht]
  \centering
  \includegraphics[width=\linewidth]{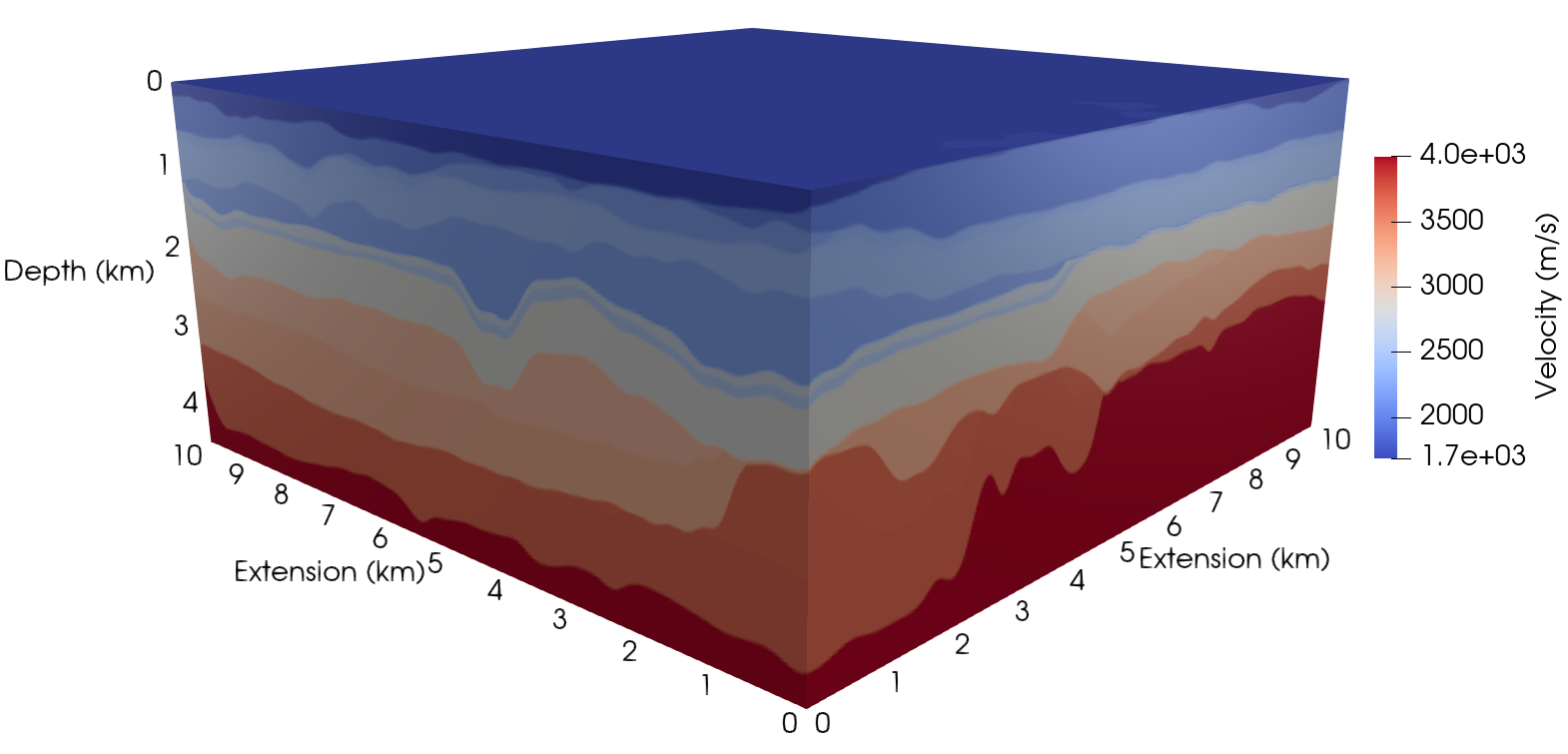}
  \caption{3-D velocity field provided by the HPC4E Seismic Test Suite.}
  \label{fig:velocity-hpc4e}
\end{figure}

In this section, we present the performance analysis of the seismic modeling and RTM using three different computational platforms: a CPU cluster, a CPU-GPU cluster, and a vector processor. The CPU cluster and CPU-GPU cluster are multicore machines from the Santos Dumont system at the National Scientific Computing Laboratory at Petrópolis/Brazil \footnote{https://sdumont.lncc.br/support\_manual.php?pg=support}. The CPU cluster has Intel Xeon E5-2695v2 Ivy Bridge processors with 2.4GHZ and 24 cores per node. On the other hand, the CPU-GPU cluster has a CPU Intel Skylake GOLD 6148, 2.4GHZ with 24 cores and 4 $\times$ NVIDIA Volta V100 per node. Finally, the vector processor is the NEC SX-Aurora TSUBASA Type 10B with 8 vector cores, and VE memory of 48GB \footnote{https://www.hpc.nec/documents/guide/pdfs/Aurora\_ISA\_guide.pdf}. To show the results for the specified platforms,  sections \ref{sec.seis_modeling}, and \ref{sec.seis_modeling3D} present the performance analysis of the seismic modeling for different grid sizes and sections \ref{sec.rtm}, and \ref{sec.rtm3D} show the analysis for the RTM for only one chosen grid size. As test cases, we have chosen the 2-D Marmousi velocity model \citep{versteeg1994marmousi} shown in Figure \ref{fig:marmousi_bench} for the 2-D experiments and the velocity field provided by the HPC4E Seismic Test Suite \footnote{https://hpc4e.bsc.es/downloads/hpc-geophysical-simulation-test-suite} for the 3-D experiments (Figure \ref{fig:velocity-hpc4e}).

\subsection{2-D Experiments}

\subsubsection{Seismic Modeling:} \label{sec.seis_modeling}

the 2-D Marmousi benchmark provides a velocity field that has a depth of 3.0 km and 9.3 km in the horizontal direction. Its original grid has $737 \times 240$ grid points, where the grid space has 12.5 meters. Considering the size thickness $N_{a} = 50$ except at the surface, where we set half of the finite difference stencil length to simulate the free-surface, the new grid size has $837 \times 294$ grid points. We create more three grids for the velocity field starting from the original one. The grids have $1574 \times 534$, $3048 \times 1014$, and $5996 \times 1974$ grid points. We use them to simulate seismic wave propagation (seismic modeling) and measure computational performance for different architectures. Each seismic modeling simulates a fixed-spread acquisition providing a single shot (seismogram) outcome. Thus, the simulation propagates the wavefield for 4 seconds with a time step of 0.5 milliseconds. We use the Ricker seismic source \citep{wang2015frequencies} of the cutoff frequency of 40 Hz that is placed near the surface.

We measure the absolute time for each run changing only the grid size. The OpenACC implementation of the seismic modeling for the  NVIDIA Volta V100 is based on Algorithm \ref{alg.seismicModelingGPU}. The vector processor implementation of the seismic modeling for the SX-Aurora TSUBASA is based on Algorithm \ref{alg.seisModeling} supported by the compilation flags presented in subsection \ref{sec.vectorImple}. We ran each application ten times, and we took the time measurements for the NVIDIA Volta V100 and SX-Aurora TSUBASA vector engine platforms that are shown in Table \ref{tab:2dseis_measurements_gpu-nec}.
We can observe in Table \ref{tab:2dseis_measurements_gpu-nec} that SX-Aurora TSUBASA performs better for all grids except for Grid 4. Besides, we observe fewer system fluctuations for the vector processor runs. On the other hand, the seismic modeling performed better on NVIDIA Volta V100 than SX-Aurora TSUBASA for the $5996 \times 1974$ grid. However, execution times can be considered of the same order because of the system fluctuations.

\begin{table*} \centering
  \caption{Seismic Modeling performance measurements for different grid sizes on the NVIDIA Volta V100 and SX-Aurora TSUBASA. The average time is calculated for ten execution time measurements.}
  \label{tab:2dseis_measurements_gpu-nec}
  \begin{tabular}{ccccc}
    \toprule
    & \multicolumn{2}{c}{NVIDIA Volta V100} & \multicolumn{2}{c}{SX-Aurora TSUBASA} \\
    \midrule
     & Average Time (s) & Variance (s) & Average Time (s) & Variance (s) \\
    \midrule
    Grid 1: 837 $\times$ 294  & 1.132 & 0.204  & 0.488 & 0.002   \\
    Grid 2: 1574 $\times$ 534  & 1.875 & 1.054  & 0.912 & 0.002   \\
    Grid 3: 3048 $\times$ 1014  & 2.615 & 0.461  & 2.416 &  0.010   \\
    Grid 4: 5996 $\times$ 1974  &  6.962 & 0.693  & 7.885 & 0.005   \\
    \bottomrule
  \end{tabular}
\end{table*}

\begin{figure}[ht]
  \centering
  \includegraphics[width=\linewidth]{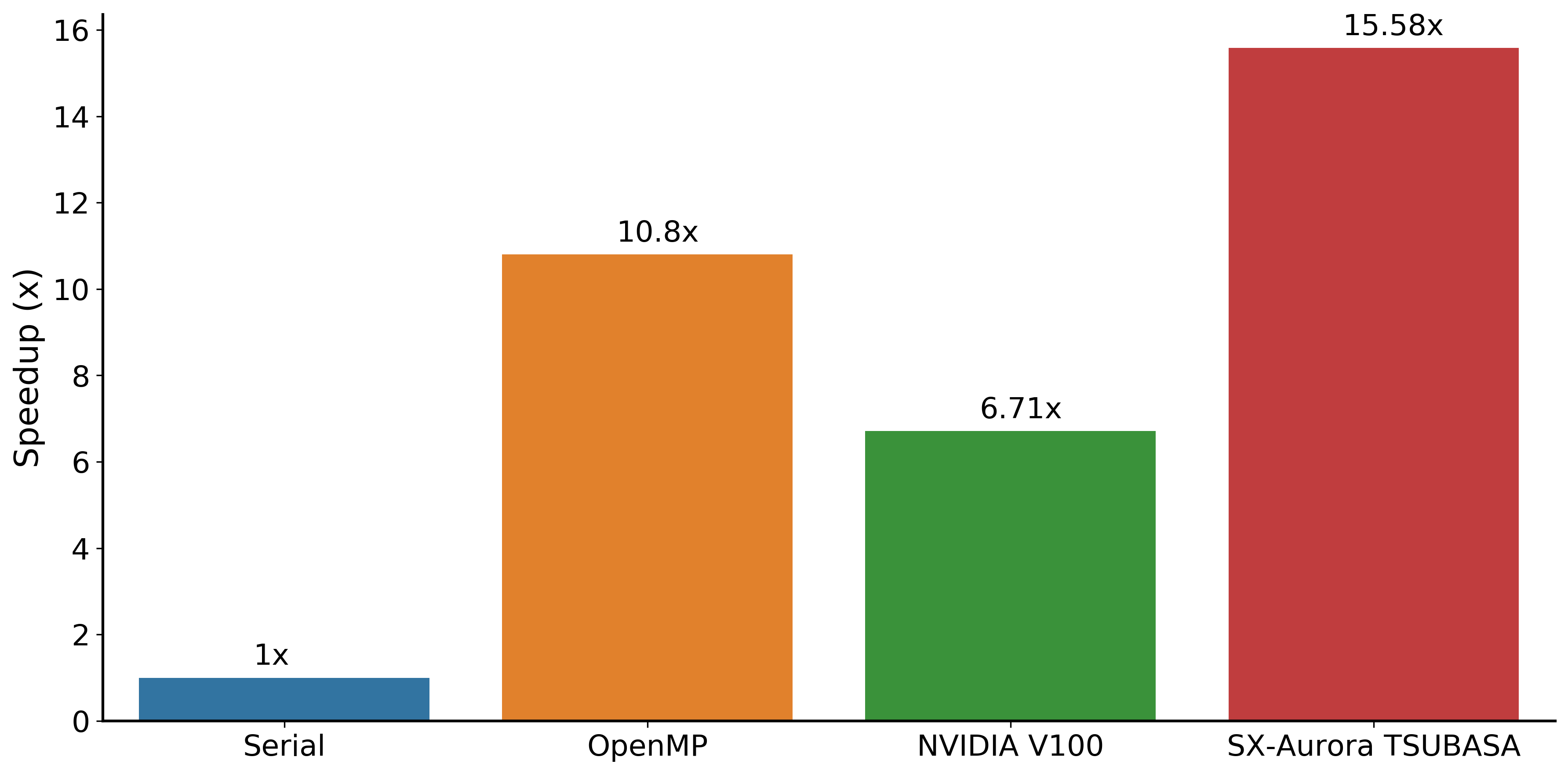}
  \caption{Seismic modeling speedup across the platforms Santos Dumont CPU Cluster, NVIDIA Volta V100 and SX-Aurora TSUBASA Vector Engine for the $837 \times 294$ grid.}
  \label{fig.speedup_modeling_grid1}
\end{figure}

\begin{figure}[ht]
  \centering
  \includegraphics[width=\linewidth]{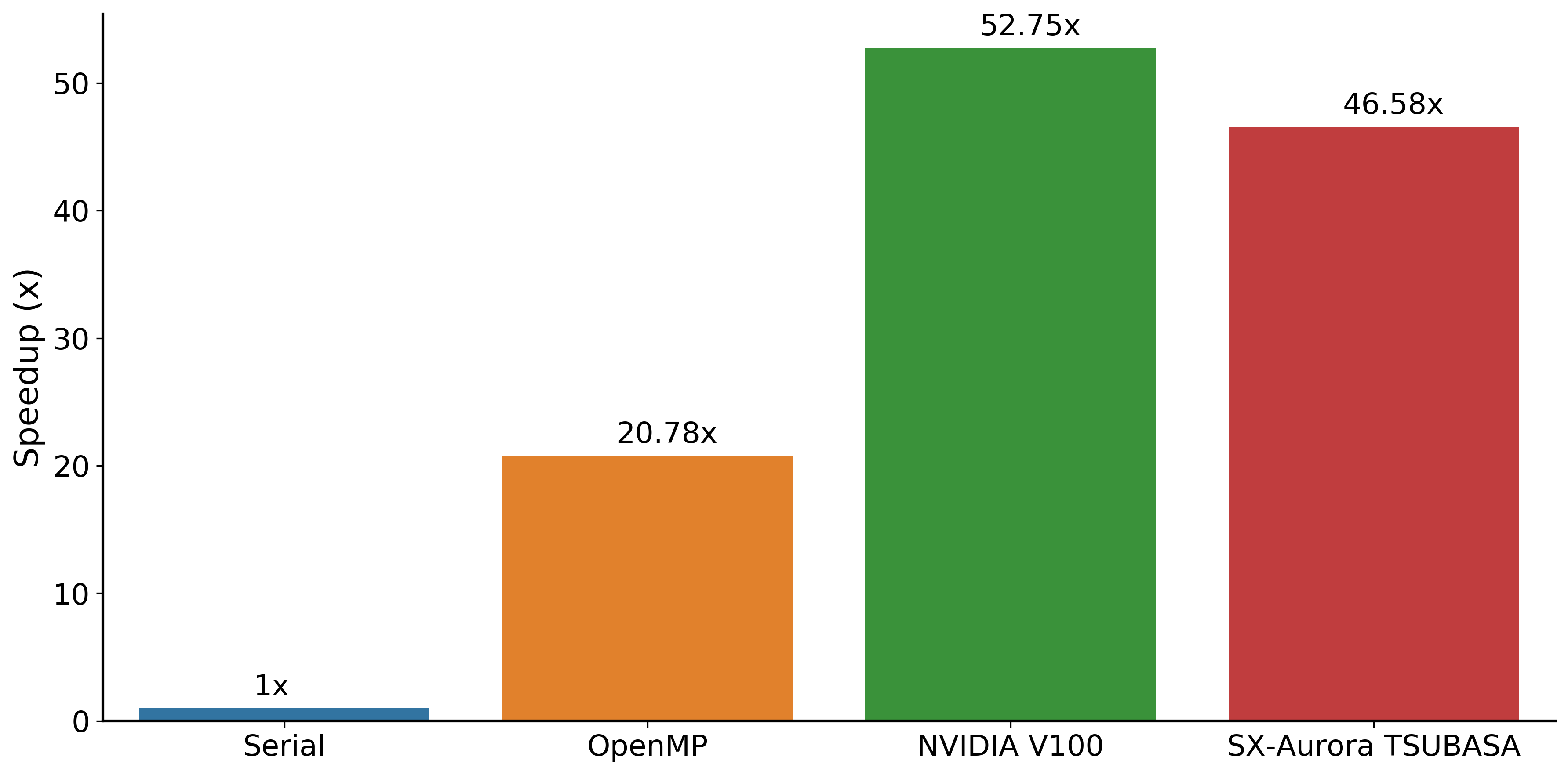}
  \caption{Seismic modeling speedup across the platforms Santos Dumont CPU Cluster, NVIDIA Volta V100 and SX-Aurora TSUBASA Vector Engine for the $5996 \times 1974$ grid.}
  \label{fig.speedup_modeling_grid4}
\end{figure}

Figures \ref{fig.speedup_modeling_grid1} and \ref{fig.speedup_modeling_grid4} show the seismic modeling speedup across the platforms Santos Dumont CPU Cluster, NVIDIA Volta V100 and SX-Aurora TSUBASA vector processor for the $837 \times 294$ and $5996 \times 1974$ grids. We ran the optimized serial and OpenMP implementations of seismic modeling on the Santos Dumont CPU Cluster. The optimized serial implementation ran on a single core, and the OpenMP version ran on 24 cores on a single node. We have chosen the execution time of the optimized serial version as the reference time to calculate the speedup. Thus, the seismic modeling speedup for the optimized serial implementation is set as 1.0. Figure \ref{fig.speedup_modeling_grid1} shows that the OpenACC implementation for the NVIDIA Volta V100 platform had the worst speedup for the $837 \times 294$ grid, that is 6.7. The seismic modeling performed better on SX-Aurora TSUBASA for the same grid size, reaching the speedup value of 15.6. Notice that the OpenMP implementation is $1.6\times$ better than the OpenACC implementation, and the vector processor implementation $1.44\times$ better than the OpenMP implementation. On the other hand, the speedup conclusions drastically change for the $5996 \times 1974$ grid. The best speedup is from the OpenACC implementation for the NVIDIA Volta V100, which is 52.8. The speedup of the vector processor implementation on SX-Aurora TSUBASA is 46.8, and the speedup of the OpenMP implementation is 20.8. Thus, the OpenACC and vector processor implementation are $2.54\times$, and $2.24\times$ better than the OpenMP implementation. Remember that SX-Aurora TSUBASA has 8 vector cores against 24 CPU cores of Santos Dumont CPU cluster.

\begin{figure}[ht]
  \centering
  \includegraphics[scale=.35]{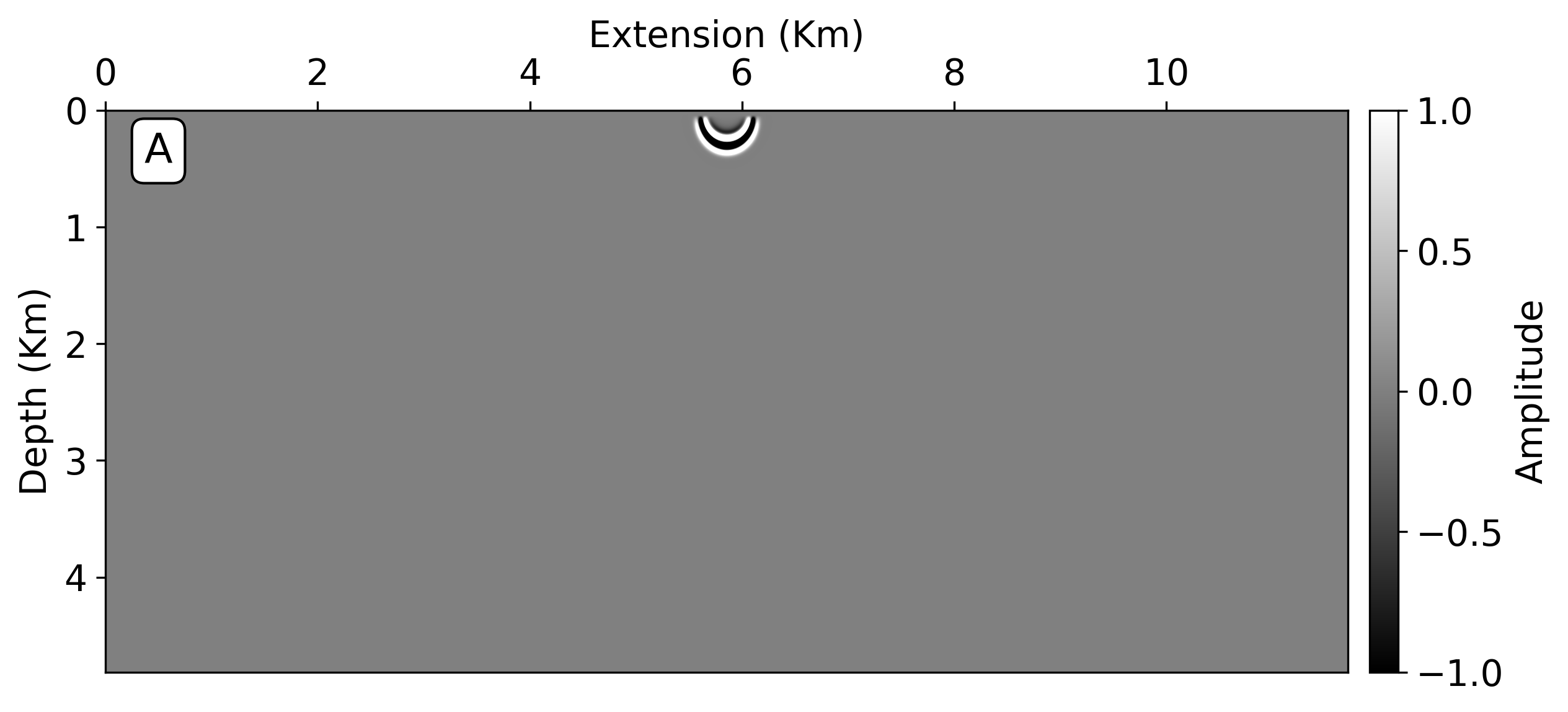}
  \includegraphics[scale=.35]{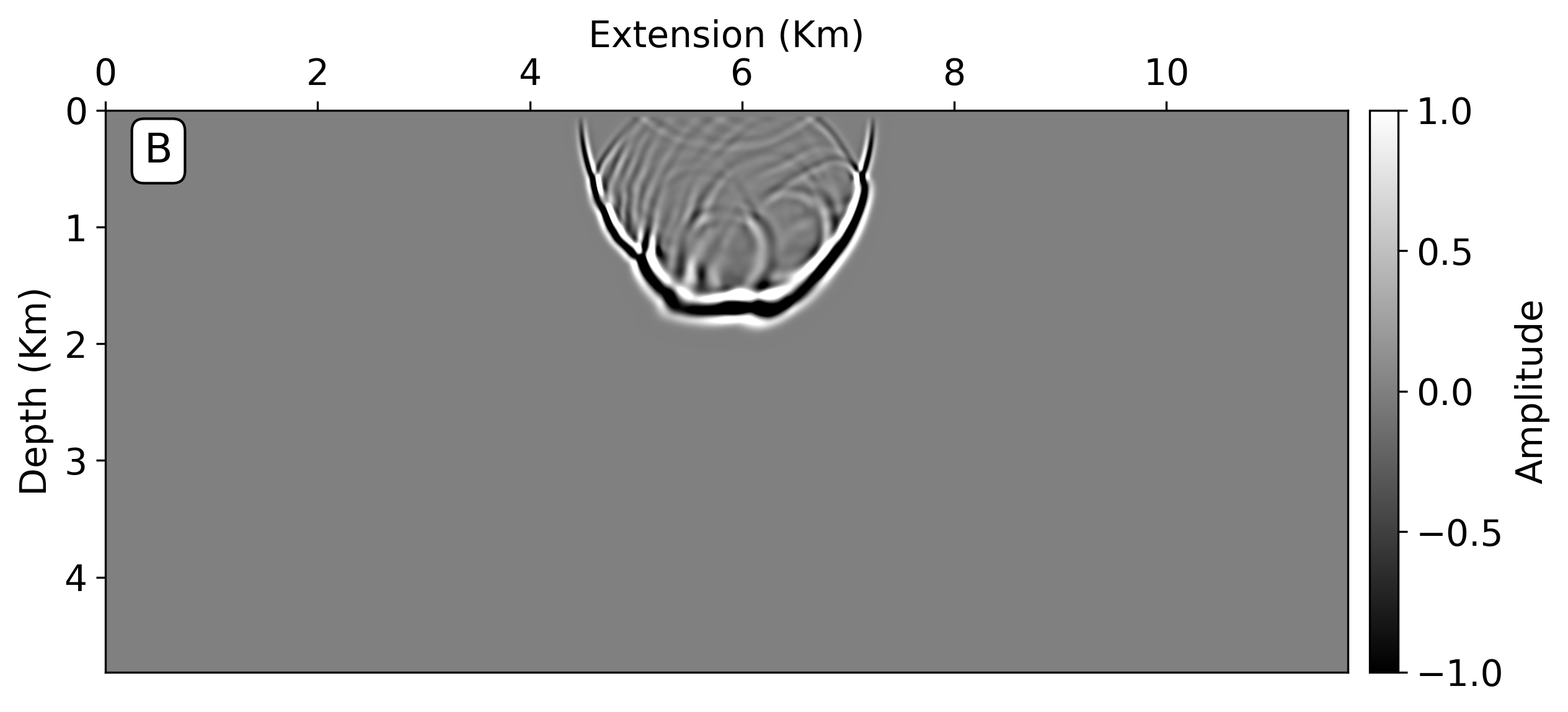}
  \includegraphics[scale=.35]{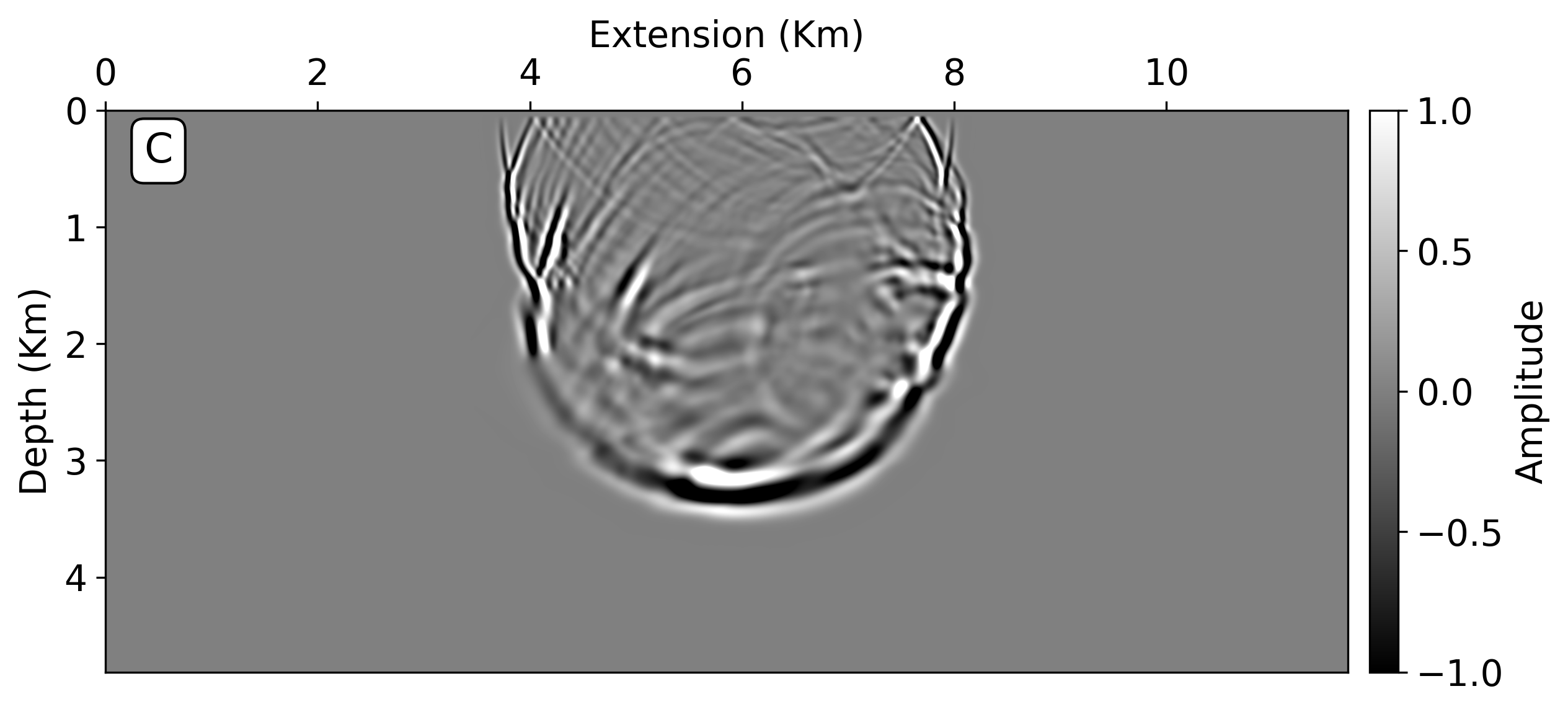}
  \includegraphics[scale=.35]{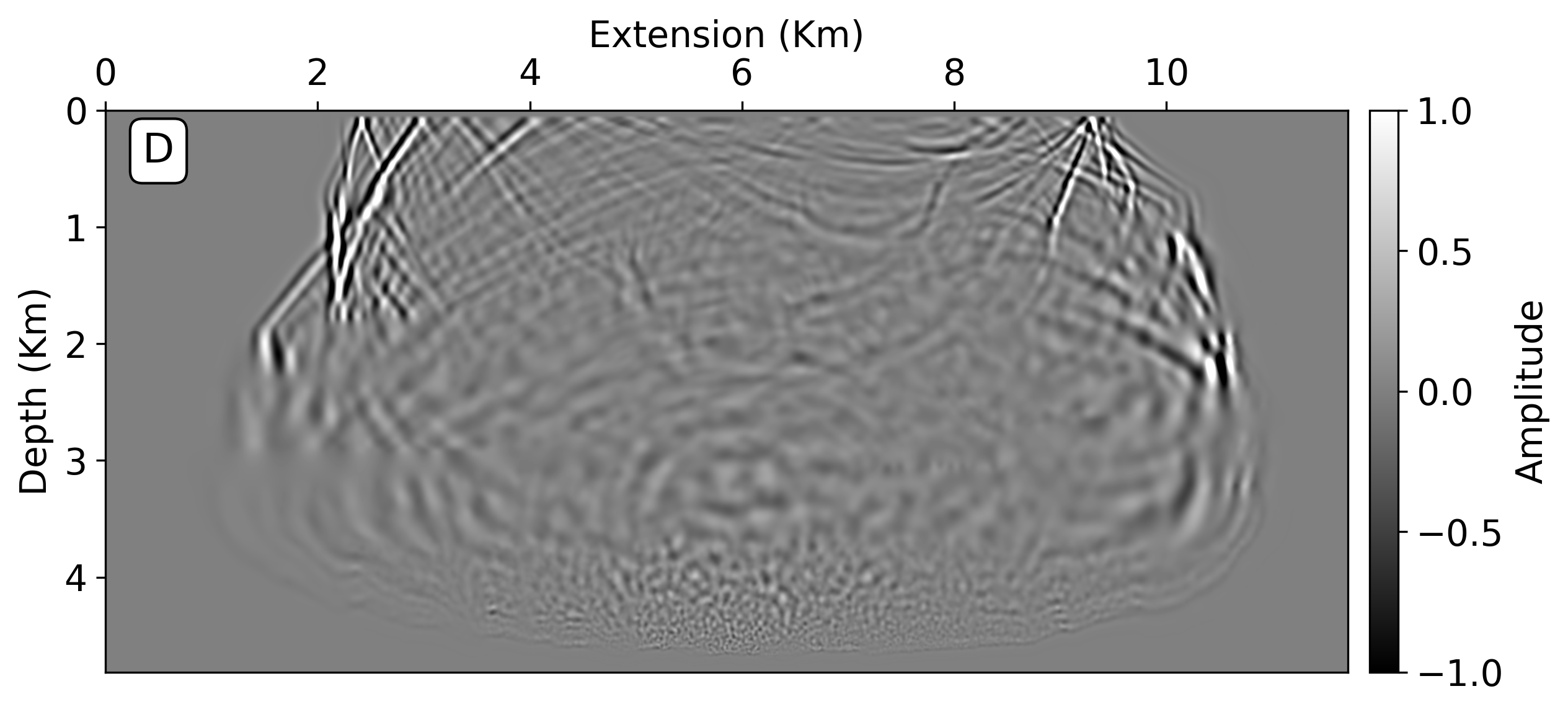}
  \includegraphics[scale=.35]{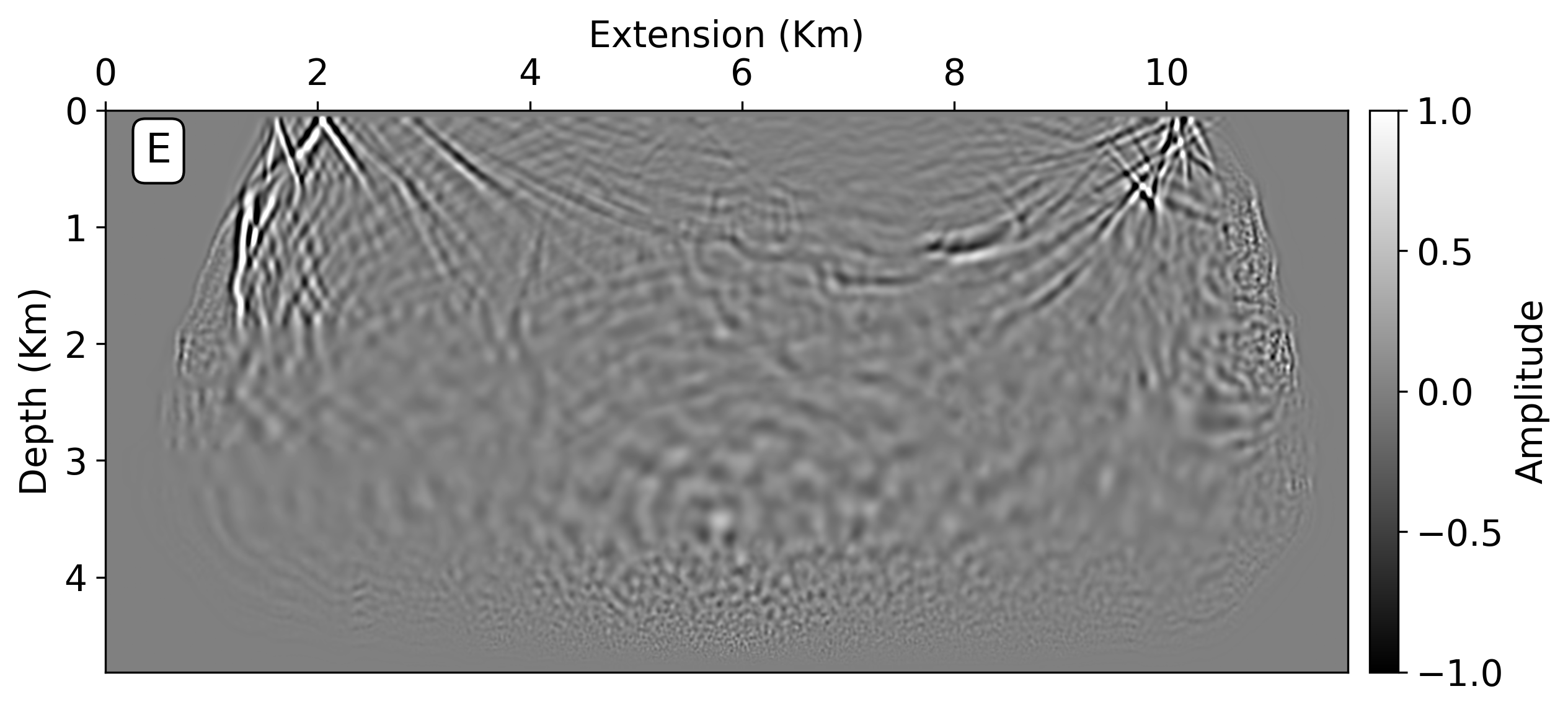}
  \includegraphics[scale=.35]{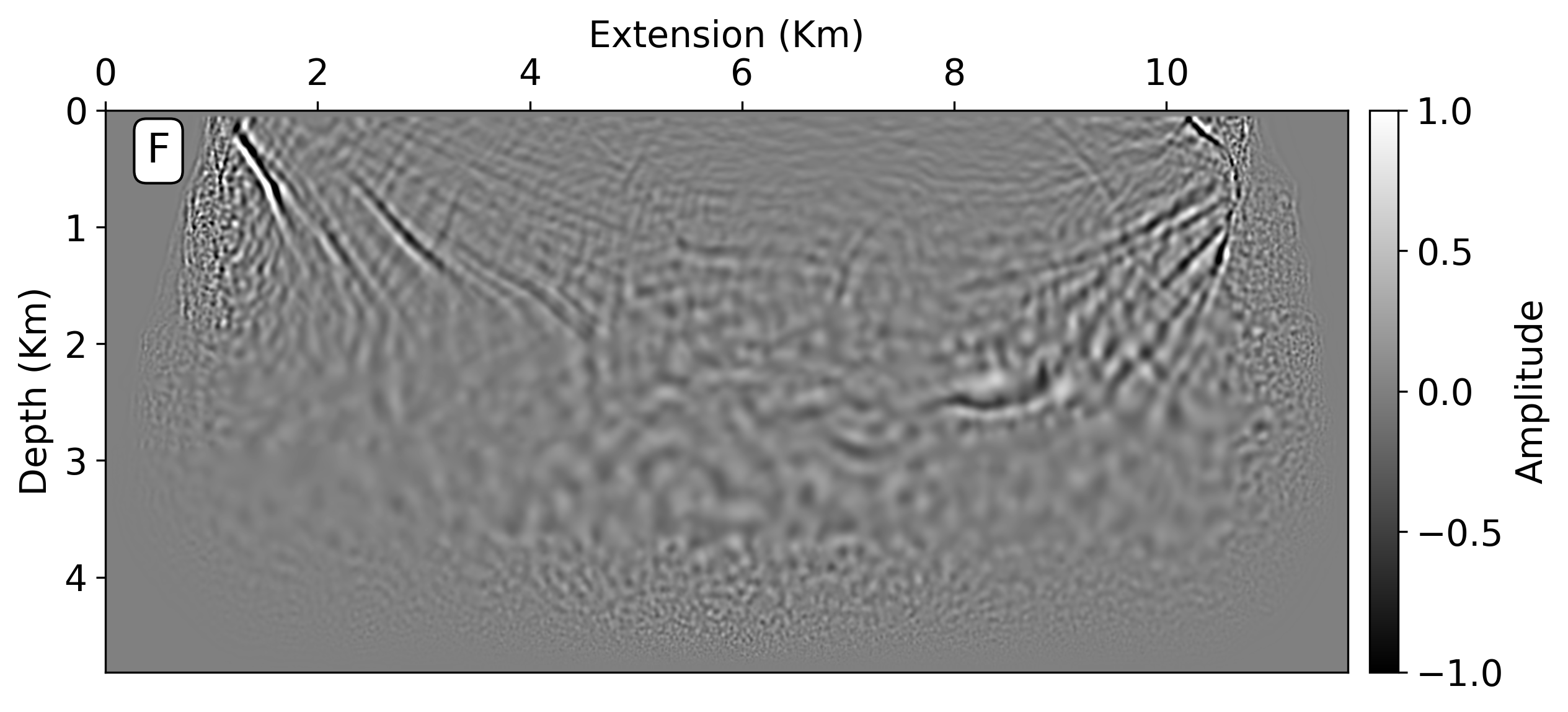}
  \caption{Propagation of the wavefield in the Marmousi velocity field with RBCs for the instants 0.25 s (A), 1.0 s (B), 1.5 s (C), 2.5 s (D), 3.0 s (E), and 3.5 s (F). The results were provided by Algorithm \ref{alg.seisModeling}.}
  \label{fig:propagation_marmousi_rbc}
\end{figure}

Lastly, Figure \ref{fig:propagation_marmousi_rbc} shows the wavefield propagation for a single shot located at $[x, z] = [4600.9, 12.5]$ meters in the Marmousi velocity field for the instants 0.25 s, 1.0 s, 1.5 s, 2.5 s, 3.0 s, and 3.5 s. Remembering, we use the Ricker seismic source \citep{wang2015frequencies} of the cutoff frequency of 40 Hz. Besides, we consider the RBC in the boundaries for the wave propagation numerical simulation, aiming to eliminate the coherent reflections. In this experiment, we use the computational implementation of Algorithm \ref{alg.seisModeling} to generate the outcomes shown in Figure \ref{fig:propagation_marmousi_rbc}.

\subsubsection{Reverse Time Migration} \label{sec.rtm}

We also use the 2-D Marmousi benchmark for the RTM experiments. However, we have been using only one grid size, which is $1574 \times 534$. The grid size considers $N_{a} = 50$ except at the top of the velocity model, where we set half of the finite difference stencil length to simulate the free surface. RTM experiment simulates a fixed-spread acquisition for one single shot. The seismic source is the Ricker wavelet \citep{wang2015frequencies} of a cutoff frequency of 40 Hz placed near the surface. We generated the observed seismogram by modeling the wave propagation and recording the signal near the surface. The tests run the RTM application and calculate the average execution time for ten measurements on the SX-Aurora TSUBASA and NVIDIA Volta V100 for further comparison. Each algorithm is supported by the OpenACC and vector processor implementations presented in subsections \ref{sec.vectorImple} and \ref{sec.openACCImple}.

Table \ref{tab:2dRTM_diskrequirement} shows the average execution time and the hard disk requirements for the 2-D RTM implementations based on Algorithm \ref{alg.rtmuq}, \ref{alg.rtmwr}, \ref{alg.rtmAlg2GPU}, and \ref{alg.rtmAlg3GPU}. Algorithms \ref{alg.rtmuq} and \ref{alg.rtmAlg2GPU} require the full storage of the source wavefield.  Nevertheless, instead of storing the source wavefield for every time step ($\Delta t$) based on the FDM, we took advantage of the Nyquist theory as explored by \citet{sun2013two} to store the wavefield at the Nyquist time step to reduce the amount of information. The Nyquist time step $\Delta t_{nyq}$ is defined as,

\begin{equation} \label{eq.nyquist_relation}
\Delta t_{nyq} = \frac{1}{2(f_{max} - f_{min})},
\end{equation}
where $f_{max}$ and $f_{min}$ are the highest and lowest frequency of the seismic source. For the Ricker wavelet that we use for the RTM test case, $f_{max} = 100.0$ Hz and $f_{min} = 0.0$ Hz. Thus, the Nyquist time step is $\Delta t_{nyq} = 5.0$ ms against to $\Delta t = 0.5$ ms for the finite difference time step.

Therefore, Algorithms \ref{alg.rtmuq} and \ref{alg.rtmAlg2GPU} which implements the wavefield storage requires $2.638$ GB of hard disk on the SX-Aurora TSUBASA and NVIDIA Volta V100 against $0.005$ GB for the wavefield reconstruction implementations (Algorithms \ref{alg.rtmwr} and \ref{alg.rtmAlg3GPU}). This represents $527.6 \times$ less information to be stored. Even using high efficient data compressors, such as the ZFP library \citep{Lindstrometal2016}, that level of  storage savings can not be achieved. For instance, \citet{barbosaenhancing} showed that using the ZFP lossy compression with the tolerance of $10^{-6}$ demands $18.64\times$ less information than the original data to be stored, which is far away from the hard disk requirements of the Algorithms \ref{alg.rtmwr} and \ref{alg.rtmAlg3GPU} implementations.

According to the measurements shown in Table \ref{tab:2dRTM_diskrequirement}, the OpenACC implementation with wavefield reconstruction presents the best time execution on average, that is 3.5 s against 4.489 s of the vector processor implementation for Algorithm \ref{alg.rtmwr}, $7.969$ s of the OpenACC implementation for Algorithm \ref{alg.rtmAlg2GPU}, and $9.016$ s of the vector processor implementation for Algorithm \ref{alg.rtmuq}. Consequently, the RTM with wavefield reconstruction is $2.28\times$ faster than the wavefield storage implementation for the NVIDIA Volta V100 and $2.01\times$ faster than wavefield storage implementation for the SX-Aurora TSUBASA. Remember that Algorithms \ref{alg.rtmwr} and \ref{alg.rtmAlg3GPU} implement one extra wave equation to reconstruct the source wavefield. 

\begin{table*} \centering
  \caption{Comparison of hard disk and time requirements for the 2-D RTM implementation with the wavefield storage, and the wavefield reconstruction.}
  \label{tab:2dRTM_diskrequirement}
  \begin{tabular}{cccc}
    \toprule
    \multicolumn{1}{c}{Method} & \multicolumn{1}{c}{Platform} & \multicolumn{1}{c}{Hard disk (GB)} & \multicolumn{1}{c}{Av. Time (s)[Variance (s)]} \\
    \toprule
    Wavefield Storage           & NVIDIA V100       & 2.638  & 7.969 [0.149] \\
    Wavefield Reconstruction    & NVIDIA V100       & 0.005  & 3.500 [1.415] \\
    Wavefield Storage           & SX-Aurora TSUBASA & 2.638  & 9.016 [0.068] \\
    Wavefield Reconstruction    & SX-Aurora TSUBASA & 0.005  & 4.489 [0.002] \\
    \bottomrule
  \end{tabular}
\end{table*}

\begin{figure}[ht]
  \centering
  \includegraphics[width=\linewidth]{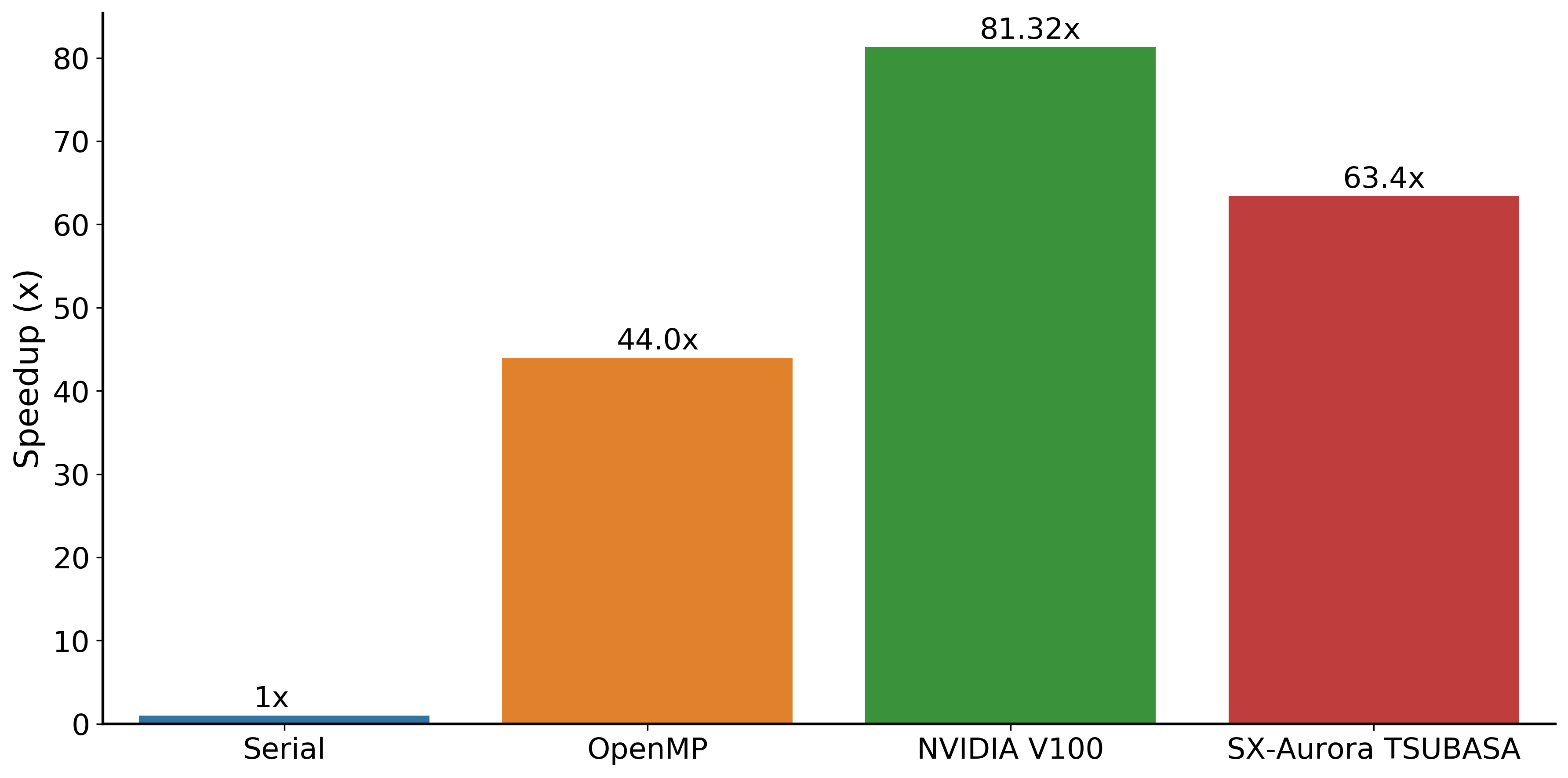}
  \caption{Reverse Time Migration speedup across the platforms Santos Dumont CPU Cluster, NVIDIA Volta V100 and SX-Aurora TSUBASA Vector Engine for the $1574 \times 534$ grid.}
  \label{fig.speedup_rtm_grid2}
\end{figure}

Concerning the speedup calculations, we calculate them only for the RTM based on Algorithms \ref{alg.rtmwr} and \ref{alg.rtmAlg3GPU} that implement the wavefield reconstruction. The OpenACC implementation presents the best result as we can see in Figure \ref{fig.speedup_rtm_grid2}. Again, our reference time is the optimized serial RTM code that was executed on the Santos Dumont CPU Cluster. We also ran the OpenMP implementation on the Santos Dumont CPU cluster using 24 CPU cores. Thus, the speedup of OpenMP implementation is  44.0 against 81.3 and 63.4 for the OpenACC and vector processor implementations. Therefore, the OpenACC speedup is $1.85\times$ better than the OpenMP speedup and $1.28\times$ better than the vector processor speedup. On the other hand, the vector processor speedup is $1.44 \times$ better than OpenMP speedup. Again, remember that SX-Aurora TSUBASA has 8 vector cores.

% -----  -----

\begin{figure}[ht]
  \centering
  \includegraphics[width=\linewidth]{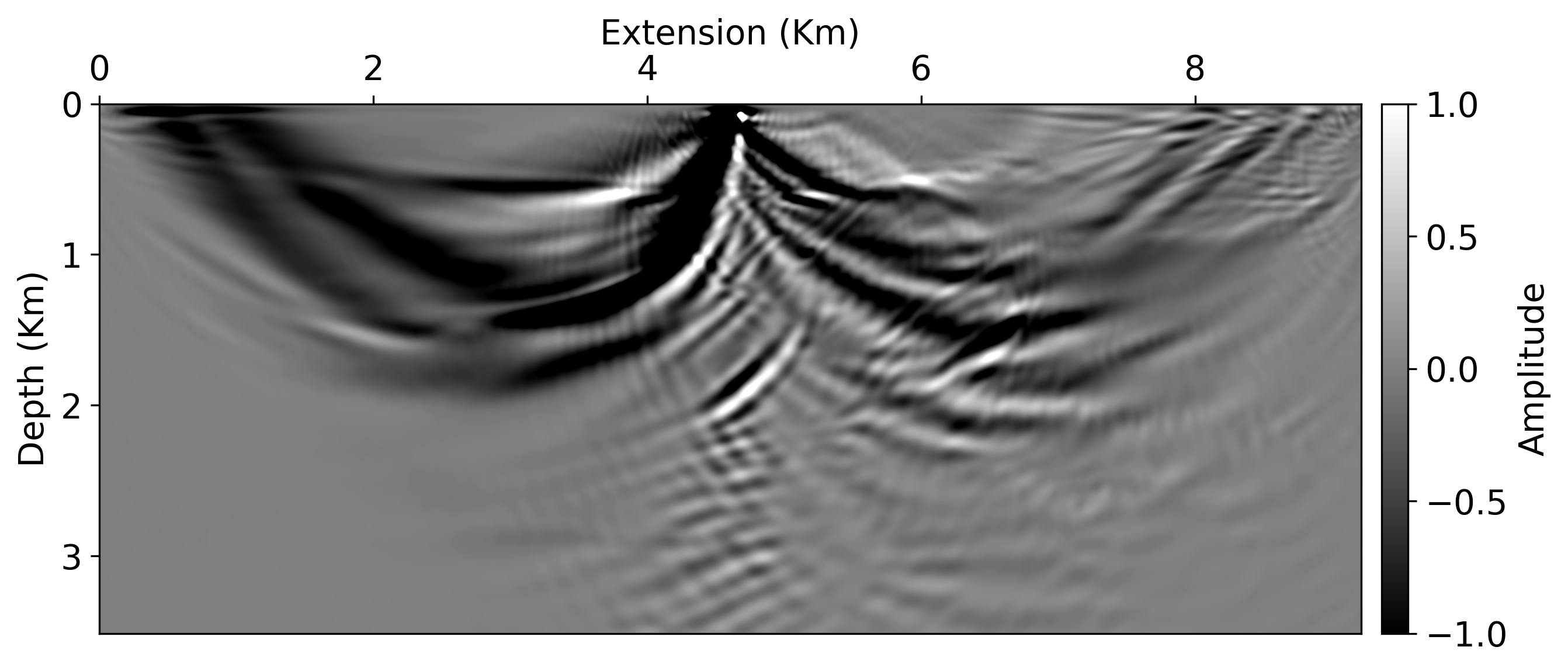}
  \caption{Seismic image for one migrated shot of 2-D Marmousi benchmark.}
  \label{fig:rtm_2d_1_shot}
\end{figure}

Figure \ref{fig:rtm_2d_1_shot} shows the partial seismic image for the 2-D Marmousi benchmark with one shot located at $[x, z] = [4600.0, 12.5]$ meters. For this experiment, we use the Ricker wavelet \citep{wang2015frequencies} of a cutoff frequency of 40 Hz and an observed seismogram generated by simulating a fixed spread acquisition recording the signals near the surface at $12.5$ meters in depth. We generated the seismic image result shown in Figure \ref{fig:rtm_2d_1_shot} based on Algorithm \ref{alg.rtmwr}, which implements the wavefield reconstruction strategy based on the IVR technique with the RBC.

\subsection{3-D Experiments}

\subsubsection{Seismic Modeling} \label{sec.seis_modeling3D}
We have chosen the MODEL AF provided by the HPC4E Seismic Test Suite for the 3-D experiments. The MODEL AF is a 3-D model designed as a set of 15 layers with constant velocity values and flat topography. Besides, the velocity parameter model covers an area of $10 \times 10 \times 4.5$ km. We have used two grid sizes for the seismic modeling experiments, where which one has $501 \times 501 \times 235$ and $901 \times 901 \times 416$ grid points with $25.0$, and $12.5$ meters of grid space, respectively. Notice that the grid sizes for the 3-D cases include $N_{a} = 50$ and half of the finite difference stencil length at the top of the velocity model to simulate the free-surface. The experiments consist of the wavefield propagation for a single shot located at $[5000, 5000]$ meters for a maximum time of $6.0$ seconds. We used the Ricker seismic source \citep{wang2015frequencies} of the cutoff frequency of 20 Hz placed near the surface. Concerning the acquisition geometry, the seismic signals (seismograms) are recorded following the expressions:

\begin{equation} \label{eq.rx_receivers}
r_{x} = 25.0 (i-1) + 1012.5 \text{ with } i = 1, \cdot \cdot \cdot, 320,
\end{equation}
\begin{equation} \label{eq.ry_receivers}
r_{y} = 25.0 (j-1) + 1012.5 \text{ with } j = 1, \cdot \cdot \cdot, 320,
\end{equation}
where, the pair $[r_{x}, r_{y}]$ meters represents the receiver locations on the surface.

We simulate the wavefield propagation based on Algorithm \ref{alg.seisModeling} for the vector processor optimizations and Algorithm \ref{alg.seismicModelingGPU} for the OpenACC implementation, both exposed in subsections  \ref{sec.vectorImple} and \ref{sec.openACCImple}. Table \ref{tab:3D_seis_measurements_gpu-nec} details the average time measurements of the seismic modeling on the NVIDIA Volta V100 and SX-Aurora TSUBASA vector engine platforms. The second and fourth columns show the time measurements for the $501 \times 501 \times 235$ grid size. The execution time is almost the same on both platforms, and it differs by approximately $1.0$ second. On the other hand, the wavefield simulation on the $901 \times 901 \times 416$ grid size took $330.363$ s for the NVIDIA VOLTA V100 and $203.904$ s for the SX-Aurora TSUBASA, representing an execution time difference of $126.423$ seconds. The results show that both platforms have similar performances for the smallest grid. However, the SX-Aurora TSUBASA performs better for the larger grid. Oppositely to the 2-D experiments, the 3-D implementations do not show relevant system fluctuations for both platforms and grid sizes in the execution time measurements.

\begin{table*} \centering
  \caption{Seismic Modeling performance measurements for different grid sizes on the NVIDIA Volta V100 and SX-Aurora TSUBASA. The average time is calculated for ten execution time measurements.}
  \label{tab:3D_seis_measurements_gpu-nec}
  \begin{tabular}{ccccc}
    \toprule
    & \multicolumn{2}{c}{NVIDIA Volta V100} & \multicolumn{2}{c}{SX-Aurora TSUBASA} \\
    \midrule
     & Average Time (s) & Variance (s) & Average Time (s) & Variance (s) \\
    \midrule
    Grid 1: 501 $\times$ 501 $\times$ 235  & 42.1505 & 0.0464  & 43.132 & 0.018   \\
    Grid 2: 901 $\times$ 901 $\times$ 416  & 330.363 & 0.0464  & 203.940 & 0.177   \\
    \bottomrule
  \end{tabular}
\end{table*}

\begin{figure}[ht]
  \centering
  \includegraphics[width=\linewidth]{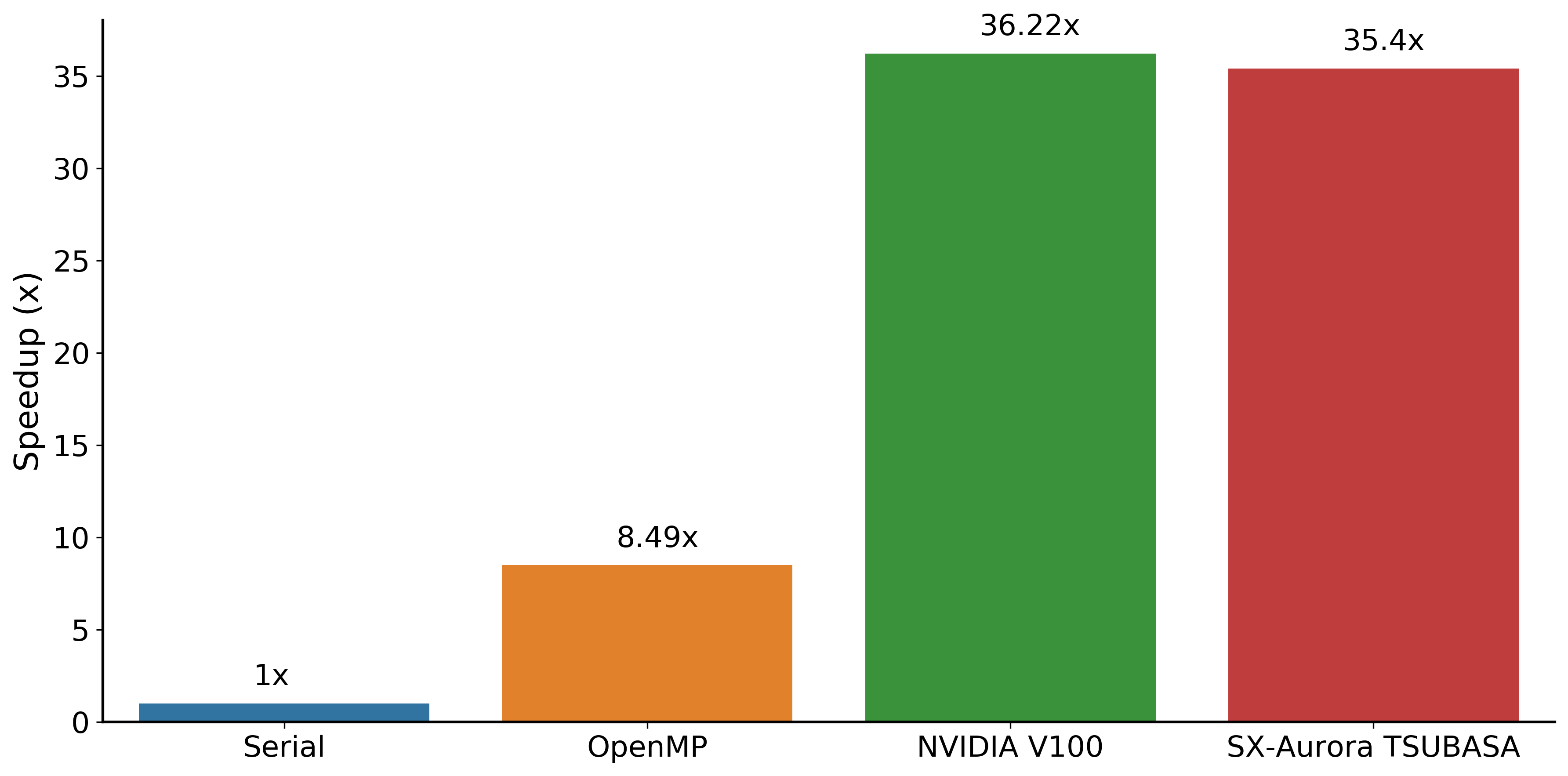}
  \caption{Seismic modeling speedup across the platforms Santos Dumont CPU Cluster, NVIDIA Volta V100 and SX-Aurora TSUBASA Vector Engine for the $501 \times 501 \times 235$ grid.}
  \label{fig.speedup_3dmodeling_grid1}
\end{figure}

\begin{figure}[ht]
  \centering
  \includegraphics[width=\linewidth]{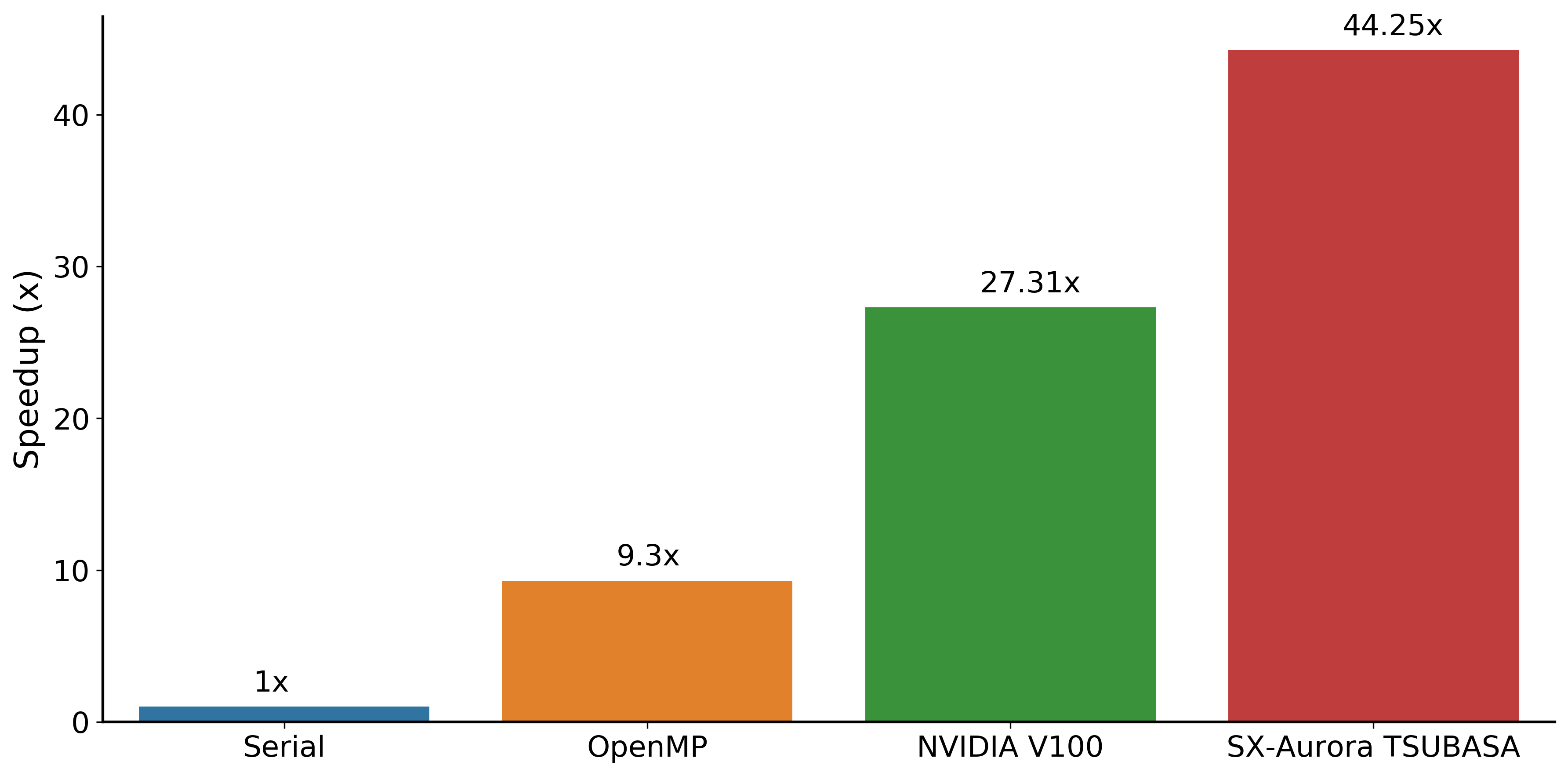}
  \caption{Seismic modeling speedup across the platforms Santos Dumont CPU Cluster, NVIDIA Volta V100 and SX-Aurora TSUBASA Vector Engine for the $901 \times 901 \times 416$ grid.}
  \label{fig.speedup_3dmodeling_grid2}
\end{figure}

Figures \ref{fig.speedup_3dmodeling_grid1} and \ref{fig.speedup_3dmodeling_grid2} show the seismic modeling speedup across the platforms Santos Dumont CPU Cluster, NVIDIA Volta V100 and SX-Aurora TSUBASA vector processor for the $501 \times 501 \times 235$ and $901 \times 901 \times 416$ grids. Again, the execution time measurement of the optimized serial version is the reference time to calculate the speedups. Thus, the seismic modeling speedup for the optimized serial implementation is set as 1.0. We ran the optimized serial, OpenMP, and OpenACC implementations of seismic modeling on the Santos Dumont CPU-GPU Cluster. The optimized serial implementation ran on a single core, and the OpenMP version ran on 24 cores on a single node. Figure \ref{fig.speedup_3dmodeling_grid1} shows that the OpenACC, and vector processor implementations have practically the same performance for the $501 \times 501 \times 235$ grid. The speedup of the seismic modeling for the OpenACC implementation on NVIDIA V100 is $36.22$, and the speedup for the vector processor implementation on SX-Aurora TSUBASA is $35.4$. Both speedups are $4.27\times$ faster than the OpenMP seismic modeling implementation. Remember that the SX-Aurora TSUBASA vector processor has 8 vector cores.

The seismic modeling implementation on SX-Aurora TSUBASA have the best performance for the $901 \times 901 \times 461$ grid size as shown in Figure \ref{fig.speedup_3dmodeling_grid2}. The speedup of the OpenMP implementation is $9.3$, the OpenACC implementation speedup is $27.31$, and the vector processor implementation speedup is $44.25$. Thus, the OpenACC implementation on NVIDIA V100 performed $1.62\times$ worse than the vector processor implementation on SX-Aurora TSUBASA and $2.94\times$ better than the OpenMP implementation. The speedup of the vector processor implementation for the SX-Aurora TSUBASA is $4.75\times$ better than the OpenMP implementation speedup.

\begin{figure}[ht]
  \centering
  \includegraphics[scale=.145]{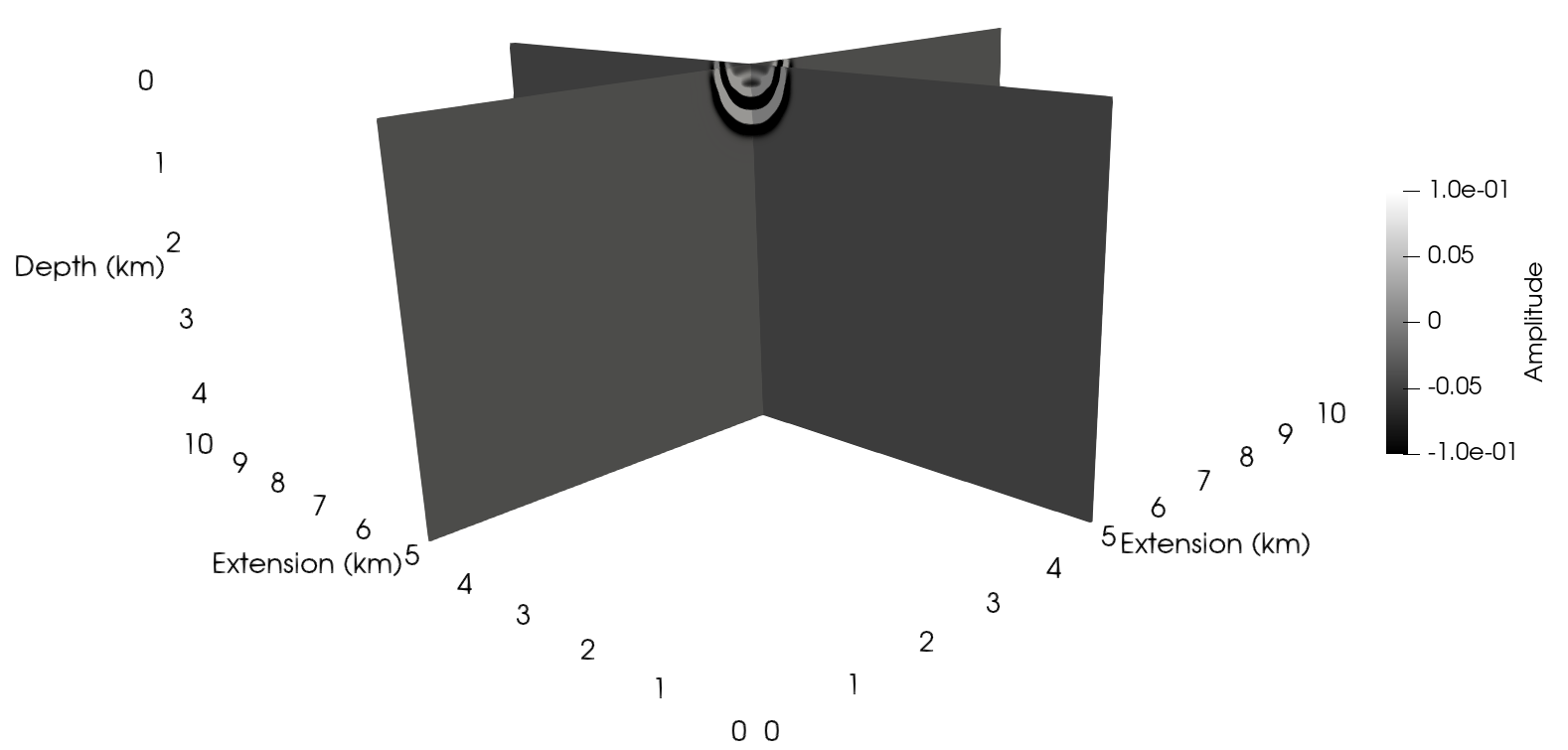}
  \includegraphics[scale=.145]{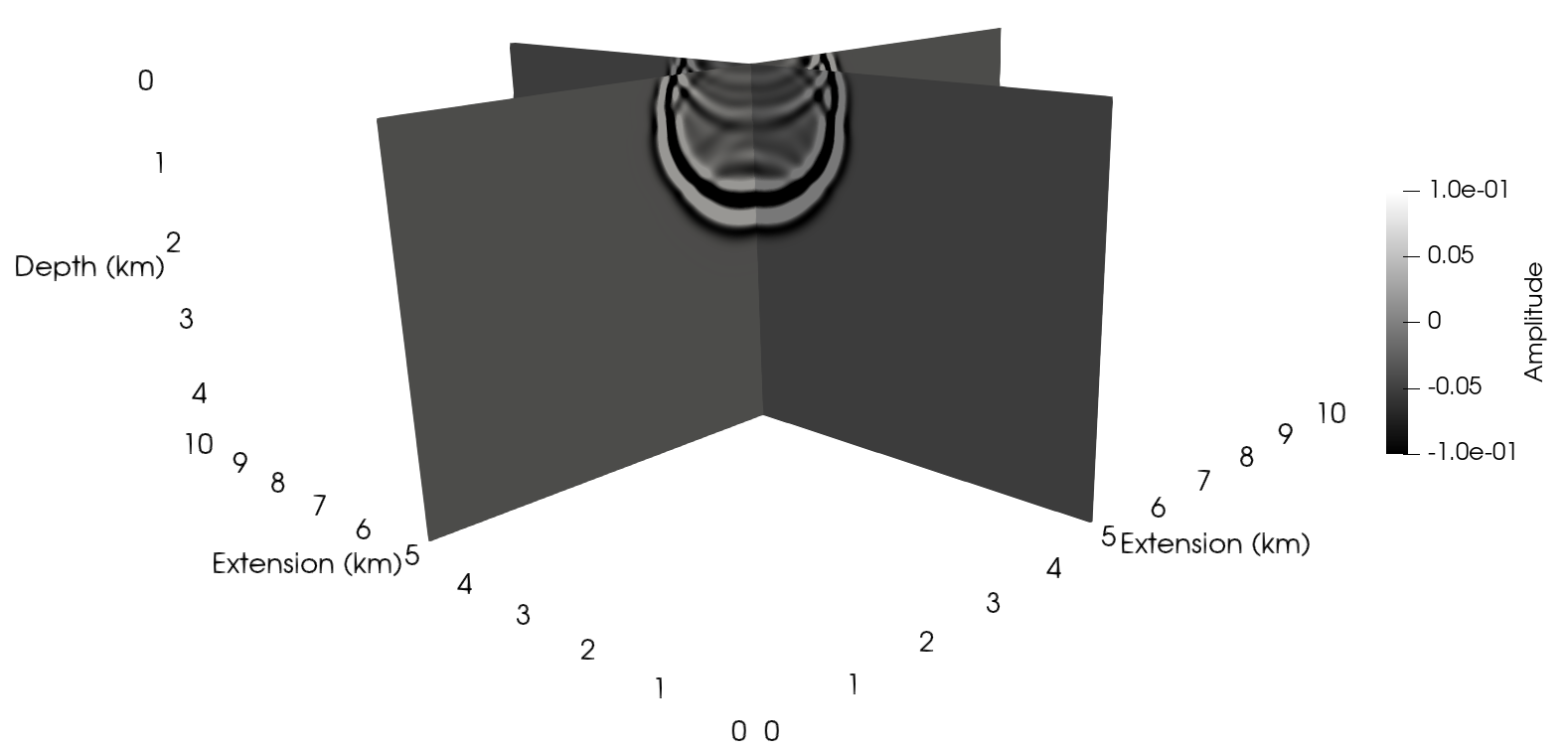}
  \includegraphics[scale=.145]{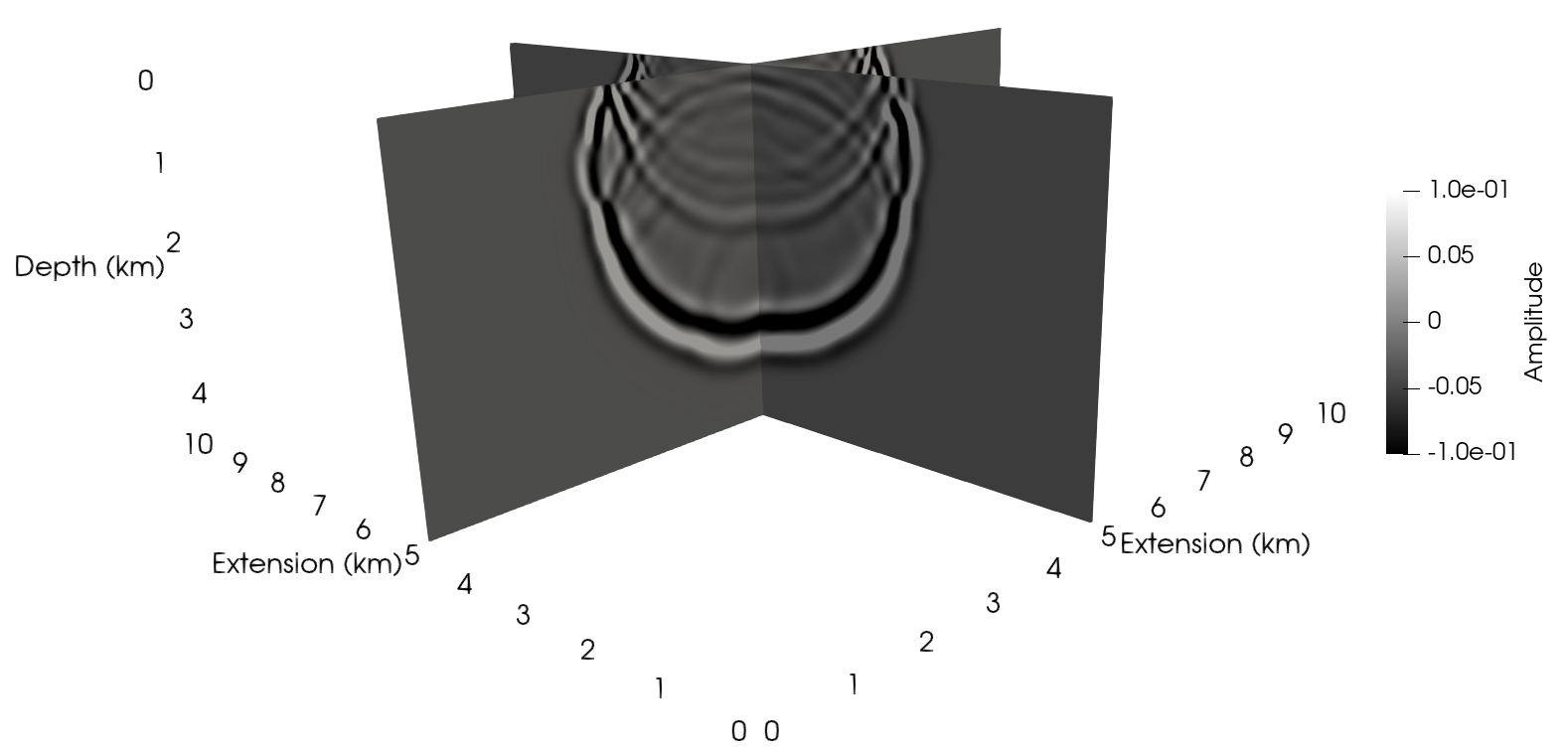}
  \includegraphics[scale=.145]{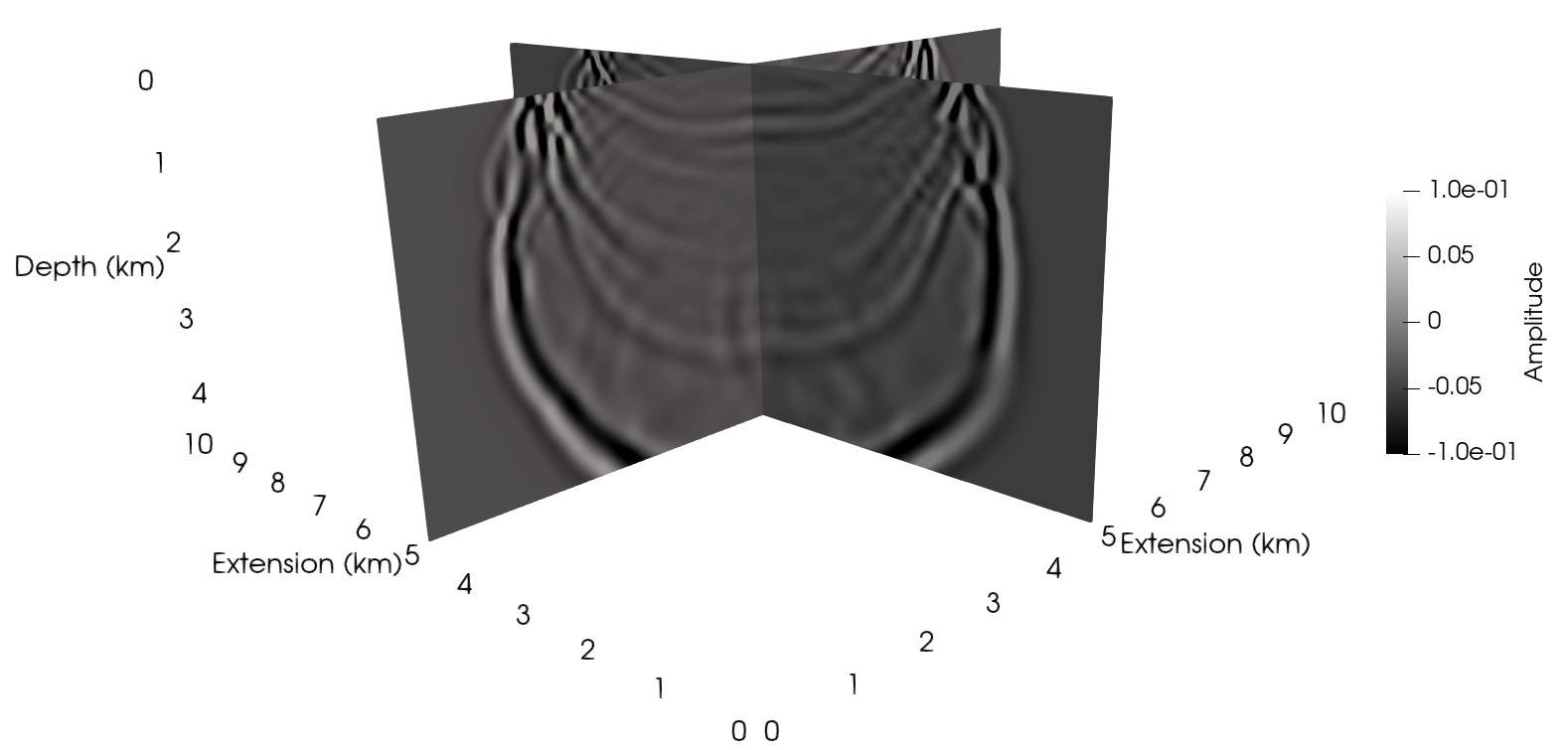}
  \caption{Propagation of the wavefield in the 3-D velocity field provided by the HPC4E Seismic Test Suite for the instants 0.6 s (upper left), 1.0 s (upper right), 1.5 s (lower left), 2.0 s (lower right). The results were provided by Algorithm \ref{alg.seisModeling}.}
  \label{fig:3d_seismic_modeling}
\end{figure}

Figure \ref{fig:3d_seismic_modeling} shows four timeframes of the wavefield propagation in the 3-D velocity field provided by the HPC4E Seismic Test Suite. The timeframes refer to the instants 0.6 s (upper left), 1.0 s (upper right), 1.5 s (lower left), 2.0 s (lower right) for a single shot located at $[x, y, z] = [5000.0, 5000.0, 25.0]$ meters. In this experiment, we ran the computational implementation of Algorithm \ref{alg.seisModeling} for the 3-D case to generate the outputs shown in Figure \ref{fig:3d_seismic_modeling}.

\subsubsection{Reverse Time Migration} \label{sec.rtm3D}

We also use the MODEL AF for the 3-D RTM experiments, but only for one grid size, which is $501 \times 501 \times 235$. The RTM simulation follows the same parameter configurations presented for the seismic modeling. Grid space of $25.0$ meters, shot location at $[5000, 5000]$ meters, maximum propagation time of $6.0$ seconds, Ricker seismic source \citep{wang2015frequencies} of cutoff frequency of $20$ Hz, and geometry acquisition following equations (\ref{eq.rx_receivers}) and (\ref{eq.ry_receivers}). We simulate the 3-D wave propagation to generate the observed seismogram by recording the signals near the surface. Again, the tests for the RTM consist of running the application based on Algorithms \ref{alg.rtmuq}, \ref{alg.rtmwr}, \ref{alg.rtmAlg2GPU}, and \ref{alg.rtmAlg3GPU} and measuring the execution time on the SX-Aurora TSUBASA and NVIDIA Volta V100 platforms.

Table \ref{tab:3dRTM_diskrequirement} shows the average time execution and the hard disk requirements for the 3-D RTM implementations for the $501 \times 501 \times 235$ grid. The RTM based on Algorithms \ref{alg.rtmuq} and \ref{alg.rtmAlg2GPU} implement the wavefield storage for the Nyquist time step described by equation (\ref{eq.nyquist_relation}), and Algorithms \ref{alg.rtmwr} and \ref{alg.rtmAlg3GPU} describe the wavefield reconstruction. We set the Nyquist time step value as $\Delta t_{nyq} = 10.0$ ms against $\Delta t = 1.0$ ms for the finite difference time step. Hence, Algorithms \ref{alg.rtmuq} and \ref{alg.rtmAlg2GPU} which implements the wavefield storage requires $132.062$ GB of hard disk for NVIDIA V100 and SX-Aurora TSUBASA against $0.439$ GB of the wavefield reconstruction implementations (Algorithms \ref{alg.rtmwr} and \ref{alg.rtmAlg3GPU}). This represents $300.82 \times$ less information to be stored.

The best average executions times refer to the RTM that implements the wavefield reconstruction on NVIDIA Volta V100 and SX-Aurora TSUBASA. Nevertheless, the vector processor implementation of the RTM performed better than the OpenACC implementation, which is $108.582$ s for the vector processor implementation against $139.786$ s for the OpenACC implementation. Considering the wavefield storage implementations in Algorithms \ref{alg.rtmuq} and \ref{alg.rtmAlg2GPU} the best average execution time is from NVIDIA Volta V100, where the OpenACC implementation took $256.166$ s to run, and the vector processor $430.944$s. The RTM with wavefield reconstruction is $1.83\times$ faster than the wavefield storage implementation for the NVIDIA Volta V100 and $3.08\times$ faster than the wavefield storage implementation for the SX-Aurora TSUBASA. The average execution time for the RTM implementation with wavefield reconstruction on SX-Aurora TSUBASA is $3.98 \times$ better than the wavefield storage on the same platform, $1.29 \times$ and $2.36$ better than the wavefield reconstruction and wavefield storage implementations for the NVIDIA Volta V100.

\begin{table*} \centering
  \caption{Comparison of hard disk and time requirements for the 3-D RTM implementation with the wavefield storage, and the wavefield reconstruction.}
  \label{tab:3dRTM_diskrequirement}
  \begin{tabular}{cccc}
    \toprule
    \multicolumn{1}{c}{Method} & \multicolumn{1}{c}{Platform} & \multicolumn{1}{c}{Hard disk (GB)} & \multicolumn{1}{c}{Av. Time (s)[Variance (s)]} \\
    \toprule
    Wavefield Storage           & NVIDIA V100       & 132.062  & 256.166 [46.810] \\
    Wavefield Reconstruction    & NVIDIA V100       &   0.439  & 139.786 [0.293] \\
    Wavefield Storage           & SX-Aurora TSUBASA & 132.062  & 430.944 [5.716] \\
    Wavefield Reconstruction    & SX-Aurora TSUBASA &   0.439  & 108.582 [0.049] \\
    \bottomrule
  \end{tabular}
\end{table*}

\begin{figure}[ht]
  \centering
  \includegraphics[width=\linewidth]{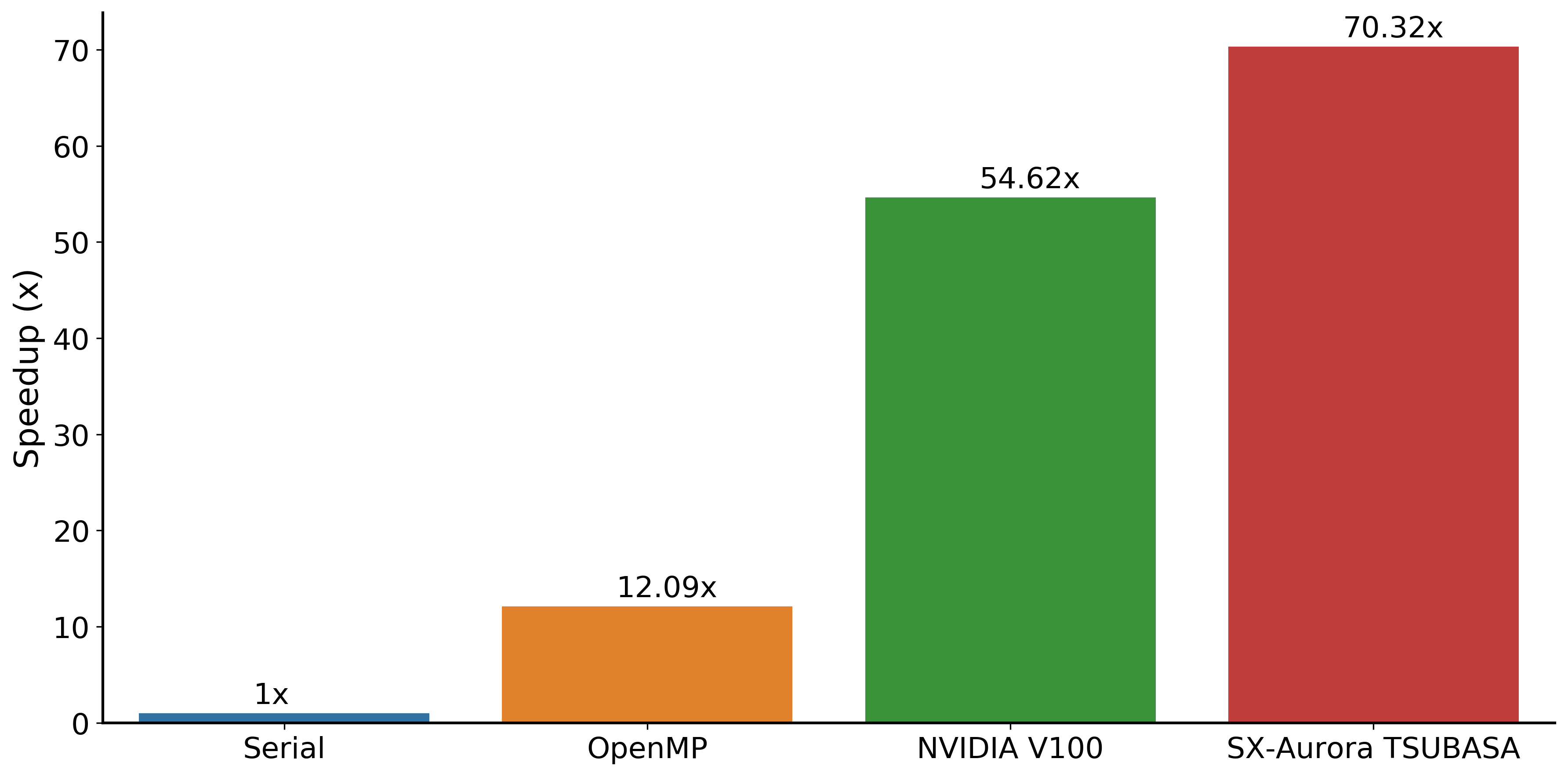}
  \caption{Reverse Time Migration speedup across the platforms Santos Dumont CPU Cluster, NVIDIA Volta V100 and SX-Aurora TSUBASA Vector Engine for the $501 \times 501 \times 235$ grid.}
  \label{fig.speedup_rtm3d_grid2}
\end{figure}

Comparing the RTM speedups across the platforms Santos Dumont CPU Cluster, NVIDIA Volta V100, and SX-Aurora TSUBASA, we can see in Figure \ref{fig.speedup_rtm3d_grid2} that the OpenMP implementation speedup is $12.09$, the OpenACC implementation speedup is $54.62$, and the vector processor implementation speedup is $70.32$. All the implementations are based on Algorithms \ref{alg.rtmwr} and \ref{alg.rtmAlg3GPU} which describe the RTM with the wavefield reconstruction. The RTM vector processor implementation has the best performance, and it is $5.82\times$ better than the OpenMP implementation and $1.28\times$ better than the OpenACC implementation. Lastly, the performance of the RTM implementation with OpenACC is $4.52\times$ OpenMP implementation.

\begin{figure}[ht]
  \centering
  \includegraphics[width=\linewidth]{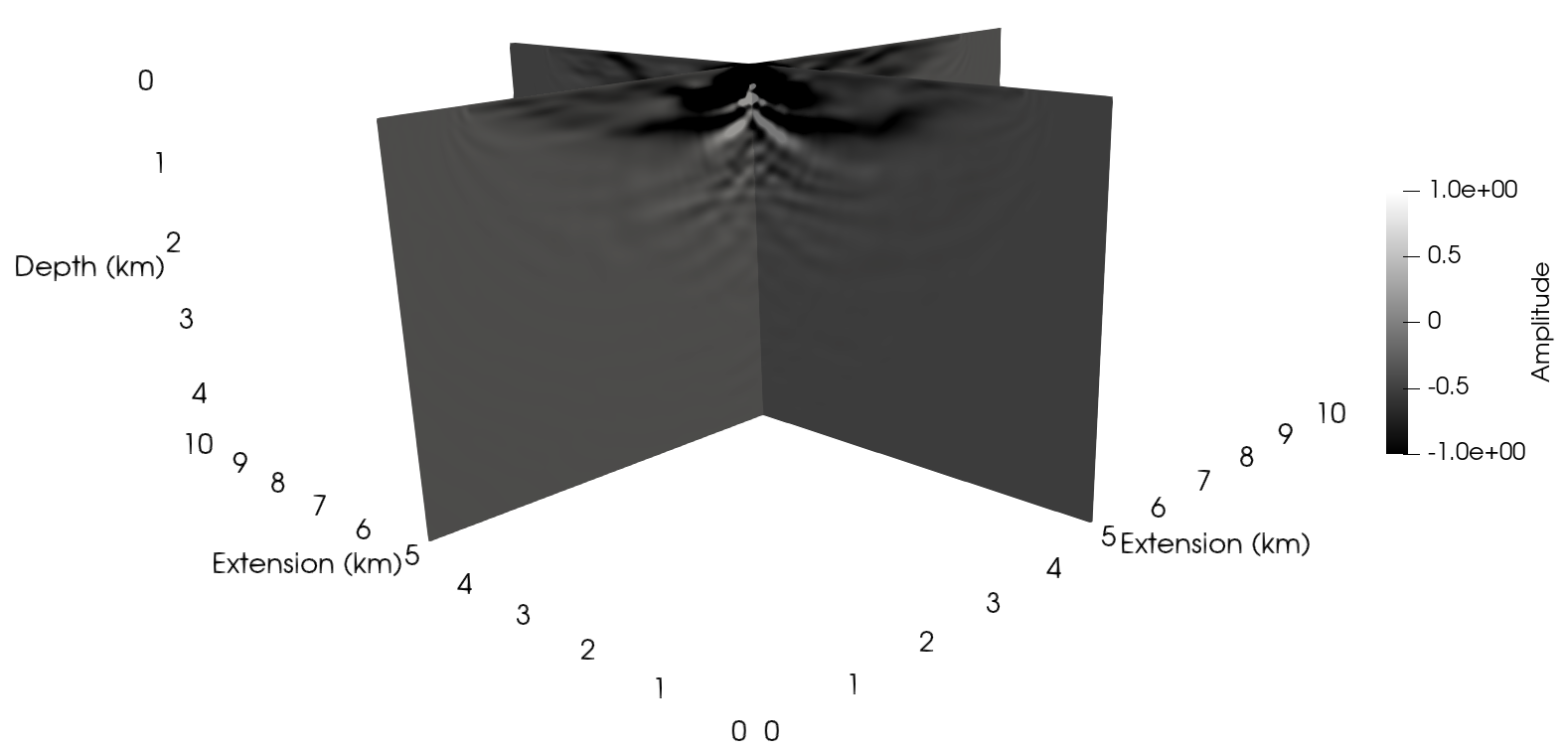}
  \caption{Seismic image for one migrated shot of 3-D HPC4E Seismic Test Suite.}
  \label{fig:rtm_3d_1_shot}
\end{figure}

Figure \ref{fig:rtm_3d_1_shot} shows the partial seismic image for one migrated shot of the 3-D HPC4E Seismic Test Suite benchmark. The shot is located at $[x, y, z] = [5000.0, 5000.0, 25.0]$ meters with the Ricker wavelet signature of a cutoff frequency of $20.0$ Hz. We generated the observed seismogram by simulating the wave propagation and recording the seismic signals at the locations following the equations (\ref{eq.rx_receivers}) and (\ref{eq.ry_receivers}) near the surface at $25.0$ meters in depth.

\section{Conclusions} \label{sec.main_findings}
This work studies RTM algorithms for 2-D and 3-D environments that mitigate the source wavefield's storage by reconstructing it through IVR based on the RBC. The RBC mitigates calculations on the artificial boundaries simplifying coding compared to versions with damping layers. Our algorithmic choices benefit computational architectures like the NEC SX-Aurora TSUBASA vector processor. For instance, our numerical experiments show that the RTM based on the wavefield reconstruction performed better on the SX-Aurora TSUBASA than on Intel Xeon multi-CPUs and NVIDIA GPU platforms for 3-D large applications. Besides, the 2-D and 3-D RTM algorithms based on the wavefield reconstruction demand less storage and are faster than the classical RTM storing the source wavefield. In this sense, the RTM based on the wavefield reconstruction demands $527.6 \times$ less information to be stored than the RTM based on wavefield storage for the 2-D test case and $300.82 \times$ less information to be stored for the 3-D test case. We also developed improved 2-D and 3-D seismic modeling algorithms with the RBC for the multi-CPU, GPU, and vector processor platforms due to their importance to LS-RTM, FWI, and UQ applications. Again, the NEC SX-Aurora TSUBASA is better for large 3D cases. We use high-level programming models such as compilation directives for the NEC SX-Aurora TSUBASA, OpenACC for the NVIDIA GPU, and OpenMP for the Multi-CPU for all computational implementations. The high-level programming models allow code portability and little code interference on the optimized baseline version. We point out that the computational implementation based on compilation flags is the simplest way to produce fast and portable codes maintaining high-performance rates. Nevertheless, further performance gains can be obtained by using tailored optimizations, sacrificing portability. 

\paragraph{Acknowledgments}
This study was financed in part by CAPES, Brazil Finance Code 001. This work is also partially supported by FAPERJ, CNPq, and Petrobras. Computer time on Santos Dumont machine at the National Scientific Computing Laboratory (LNCC - Petrópolis), and on the NEC SX-Aurora TSUBASA at NEC LATIN AMERICA S.A. is also acknowledged.

% --------------------------------------------------------------------------
\bibliographystyle{unsrtnat}
\bibliography{references}  %%% Uncomment this line and comment out the ``thebibliography'' section below to use the external .bib file (using bibtex) .

%%% Uncomment this section and comment out the \bibliography{references} line above to use inline references.
% \begin{thebibliography}{1}

% 	\bibitem{kour2014real}
% 	George Kour and Raid Saabne.
% 	\newblock Real-time segmentation of on-line handwritten arabic script.
% 	\newblock In {\em Frontiers in Handwriting Recognition (ICFHR), 2014 14th
% 			International Conference on}, pages 417--422. IEEE, 2014.

% 	\bibitem{kour2014fast}
% 	George Kour and Raid Saabne.
% 	\newblock Fast classification of handwritten on-line arabic characters.
% 	\newblock In {\em Soft Computing and Pattern Recognition (SoCPaR), 2014 6th
% 			International Conference of}, pages 312--318. IEEE, 2014.

% 	\bibitem{hadash2018estimate}
% 	Guy Hadash, Einat Kermany, Boaz Carmeli, Ofer Lavi, George Kour, and Alon
% 	Jacovi.
% 	\newblock Estimate and replace: A novel approach to integrating deep neural
% 	networks with existing applications.
% 	\newblock {\em arXiv preprint arXiv:1804.09028}, 2018.

% \end{thebibliography}

\end{document}